\documentclass[a4paper,12pt]{article}

\usepackage{times}
\usepackage{geometry}
\usepackage{float}
\usepackage{newtxmath}
\usepackage[makeroom]{cancel}
\usepackage{relsize}
\usepackage[T2A,T1]{fontenc}
\usepackage[utf8]{inputenc}
\usepackage{euscript}
\usepackage{upgreek}
\usepackage[dvipsnames]{xcolor}
\usepackage{booktabs}
\usepackage{physics}
\usepackage{makecell}

\usepackage[font={color=black},justification=centering]{caption}
\usepackage{multirow}
\usepackage[export]{adjustbox}
\usepackage{pdflscape}
\definecolor{skyblue}{HTML}{56B4E9}
\definecolor{cerulean}{HTML}{00557F}
\definecolor{ochre}{HTML}{CC7722}
\definecolor{natblue}{HTML}{0089CB}
\definecolor{bluegreen}{HTML}{0D98BA}
\definecolor{caribbeangreen}{HTML}{00CC99}
\definecolor{egyptianblue}{HTML}{1034A6}
\definecolor{myrtle}{HTML}{21421E}
\definecolor{natgreen}{HTML}{2E9F5C}
\definecolor{crimson}{HTML}{DC143C}
\definecolor{grey}{rgb}{0.5,0.5,0.5}
\definecolor{forestgreen}{HTML}{00CC99}

\usepackage[referable]{threeparttablex}
\usepackage{enumitem}
\usepackage{graphicx}
\usepackage[figuresright]{rotating}
\AtBeginDocument{\pagenumbering{arabic}}
\usepackage{array}
\newcommand{\commentout}[1]{\ignorespaces}
\usepackage[colorlinks = true,bookmarksnumbered]{hyperref}
\usepackage{amsmath,amsfonts}%
\usepackage{nicefrac}
\usepackage[nameinlink]{cleveref}
\crefname{table}{Table}{Tables}
\crefname{figure}{Fig.}{Figures}

\usepackage{lineno}
\usepackage[english]{babel}
\usepackage[autostyle]{csquotes}
\usepackage[symbol]{footmisc}
\usepackage{threeparttable}
\usepackage{fontawesome}

\hypersetup{
     colorlinks=true,
     linkcolor=crimson,
     filecolor=egyptianblue,
     citecolor = caribbeangreen,      
     urlcolor=natgreen,
     }
\newcommand{\email}[1]{\href{mailto:#1}{\textcolor{egyptianblue}{#1}}}
\newcommand{\keywords}[1]{\textbf{Keywords:} #1}

\usepackage{chngcntr}

\makeatletter
\def\@listiii{\leftmargin\leftmarginiii
              \labelwidth\leftmarginiii
              \advance\labelwidth-\labelsep
              \topsep\z@
              \parsep\z@
              \partopsep\z@
              \itemsep\topsep}
\makeatother

\DeclareSymbolFont{cyrillic}{T2A}{cmr}{m}{n}
\def\makecyrsymbol#1#2{%
  \begingroup\edef\temp{\endgroup
    \noexpand\DeclareMathSymbol{\noexpand#1}
    {\noexpand\mathalpha}{cyrillic}%
    {\expandafter\expandafter\expandafter
     \calccyr\expandafter\meaning\csname T2A\string#2\endcsname\end}}%
  \temp}
\expandafter\def\expandafter\calccyr\string\char#1\end{#1}

\makecyrsymbol\Zhe\CYRZH

\def\fontsubfuzz{9999pt}
\def\anyfontsize@set#1 #2\hfuzz{%
  \@tempdimb=\f@user@size\p@%
  \edef\external@font{#1 at\the\@tempdimb}%
}
\gdef\tryif@simple#1-#2\tryif@simple{%
  \let \reserved@f \try@simples
  \if>#2%
    \dimen@ #1\p@
    \ifdim \dimen@<\@M\p@
      \ifdim \f@size\p@<\dimen@
        \@tempdimc \dimen@     
        \advance\@tempdimc -\f@size\p@
      \else                           
        \@tempdimc \f@size\p@
        \advance\@tempdimc -\dimen@
      \fi                          
      \ifdim \@tempdimc<\@tempdimb
        \@tempdimb \@tempdimc     
        \def \best@size{#1}%
      \fi                   
    \else
  \ifx \external@font\@empty
    \ifx \best@size\@empty  
    \else                 
      \ifdim \@tempdimb>\font@submax \relax
        \xdef \font@submax {\the\@tempdimb}%
      \fi                                   
      \let \f@user@size \f@size
      \let \f@size \best@size  
      \ifdim \@tempdimb>\fontsubfuzz\relax
        \@font@warning{Font\space shape\space
            `\curr@fontshape'\space in\space size\space
             <\f@user@size>\space not\space available\MessageBreak
             size\space <\f@size>\space substituted}%
      \fi                                            
      \try@simple@size
      \expandafter\anyfontsize@set\external@font\hfuzz
      \do@subst@correction
    \fi                   
 \fi
      \let \reserved@f \remove@to@nnil
    \fi                               
  \fi
  \reserved@f}

\newcommand{\textcursive}[1]{{\fontencoding{T1}\fontfamily{frc}\selectfont #1}}

\makeatletter 
\newcommand\semismall{\@setfontsize\semismall{10.0}{12.6}}
\makeatother

\newenvironment{natabstract}{%
    \begin{center}
    \textbf{\Large Abstract} \\[1em]
    \end{center}
    \begin{quote}
 \bf}{%
    \end{quote}
}

\title{\vspace{-0.2in}Terrestrial atmospheric ion implantation occurred in the nearside lunar regolith during the history of Earth's dynamo}

\author{Shubhonkar Paramanick$^{1,\ddagger,}$\textsuperscript{\semismall{\faEnvelopeO}}, Eric G. Blackman$^{1,2,}$\textsuperscript{\semismall{\faEnvelopeO}}, John A. Tarduno$^{3,1,2,}$\textsuperscript{\semismall{\faEnvelopeO}}, \\
Jonathan Carroll-Nellenback$^{1,2,}$\textsuperscript{\semismall{\faEnvelopeO}} \\
\\
\normalsize{$^{1}$Department of Physics and Astronomy, University of Rochester, Rochester, 14627, NY, USA}\\
\normalsize{$^{2}$Laboratory for Laser Energetics, University of Rochester, 250 E River Road, Rochester, 14623, NY, USA}\\
\normalsize{$^{3}$Department of Earth and Environmental Sciences, University of Rochester, Rochester, 14627, NY, USA}\\
\\
\normalsize{$^{\ddagger}$To whom correspondence should be addressed.} \\
\\
\normalsize{\textsuperscript{\semismall{\faEnvelopeO}} E-mails: \email{shubhonkar.paramanick@rochester.edu} (S.P.), \email{eric.blackman@rochester.edu} (E.G.B.),} \\ \normalsize{\email{john.tarduno@rochester.edu} (J.A.T.), \email{jonathan.carroll@rochester.edu} (J.C.-N.)}
}

\date{}

\usepackage{fancyhdr}
\usepackage{lastpage}
\renewcommand{\footnotesize}{\fontsize{8.5}{10}\selectfont}

\fancypagestyle{plain}{
\fancyhead{}
\fancyfoot{}
\fancyfoot[L]{\color{grey}
Evidence of Earth wind implantation on Moon during geodynamo history
}
\fancyfoot[R]{\color{grey}Page \thepage\ of \pageref{LastPage}}

\renewcommand{\footnotesize}{\fontsize{8.5}{10}\selectfont} 
}

\begin{document} 


\baselineskip24pt


\maketitle 


\begin{natabstract}
Light volatile elements in lunar regolith are thought to be a mixture of the solar wind and Earth's atmosphere, the latter sourced in the absence of geomagnetic field. However, the extent to which both the current and primitive geodynamo influence the transport of terrestrial ions still remains unclear, and this uncertainty is further complicated by the enigmatic composition and poorly constrained location of the Eoarchean exosphere. Here we use 3-D MHD numerical simulations with present-day magnetized and Archean unmagnetized atmospheres to investigate how Earth's intrinsic magnetic field affects this transfer, aiming to constrain how and when the lunar isotopic signature provides a record of Earth's paleoatmosphere. We find that atmospheric transfer is efficient only when the Moon is within Earth's magnetotail. The non-solar contribution to the lunar soil is best explained by implantation during the long history of the geodynamo, rather than any short, putatively unmagnetized epoch of early Earth. This further suggests the history of the terrestrial atmosphere, spanning billions of years, could be preserved in buried lunar soils. Our results indicate that the elemental abundances of Apollo samples are very sensitive to Earth's exobase altitude, which, at the time of ion implantation, was never smaller than 190 km.
\end{natabstract}


\section*{Teaser}
Solar wind-driven transport of atmospheric ions from a magnetized Earth accounts for the non-solar component in lunar soil.

\section{Introduction}
Elements such as H, C, N, and the light noble gases in lunar soils are essentially absent from lunar rocks but appear in the lunar regolith. This dearth of volatile elements in lunar rocks has been recognized since the Apollo missions and their exclusive presence in the lunar soil requires an extra-lunar source. One extra-lunar source is commonly recognized to be the solar wind (SW), which directly implants energetic ions on the Moon's surface. The SW can account for some light volatile elements, but the overabundance of nitrogen compared to standard solar composition and the highly variable isotopic composition $\mathrm{(^{15}N \big/ ^{14}N)}$ has been a challenge to explain.

\subsection{Lunar Nitrogen Conundrum}
Possible explanations for the enrichment of N in the lunar regolith have included (i) elemental fractionation during lunar processes \cite{Kerridge1993}; (ii) N from a non-solar origin \cite{Wieler1999,Geiss1982}; and (iii) monotonic secular change in the SW due to the high-energy events near the solar surface over the last 2.5 billion years \cite{Kerridge1975,Clayton1980,Becker1975}. Lunar soils seem to contain more N than what spallation processes could have generated from incoming cosmic rays \cite{Hashizume2000,Mortimer2016}, favoring that some  lunar N is extra-solar. Proposed sources of N have included interplanetary dust \cite{Hashizume2000}, outgassing from the lunar volcanoes, fractionated terrestrial atmosphere, and intensive meteorite, cometary or asteroid bombardment shortly after formation of the Moon \cite{Hashizume2002,Marty2003,Evelyn2015}. 

Ozima et al. \cite{Ozima2005} attributed most of the extrasolar N, along with some other noble gases in lunar soil, to ion flows from Earth's atmosphere during a putative period when the geomagnetic field was absent early in Earth's history. Terada et al. \cite{Terada2017} also argued that the lunar soil oxygen isotope ratios are best explained by a contribution from Earth's atmosphere. Poppe et al. \cite{Poppe2021,Barani2024} observed terrestrial ions in the Earth's magnetotail, while Li et al. \cite{Li2023} attributed their observations of lunar surface water to the solar wind as the singular source during the Moon's passage through the magnetotail. However, these studies failed to consider how the two end members—solar wind and escaping Earth atmospheric ions—interact and mix during their transport. Besides, these analyses did not include a dynamic model of how N or O ions escaped from the atmosphere, and thus, whether the absence of a geomagnetic field is essential for the transport of such ions from Earth's atmosphere to the Moon was left unresolved.

\subsection{Does the geomagnetic field affect the atmospheric mass loss?}
The present intrinsic dipolar geomagnetic B-field interacts with the wind and forms a magnetospheric cavity extending to the SW stand-off location at $\sim 8-11$ Earth radii $(R_{\mathsmaller{\oplus}})$ on the dayside and to the distant neutral point at $X \sim 100 \, R_{\mathsmaller{\oplus}}$ on the nightside \cite{Nishida1995,Slavin1985}. There, the open field lines, which  originated from  dayside reconnection, again reconnect to form magnetospheric closed field lines (Figure 1). Lammer et al. \cite{Lammer2009} argued that such a configuration helps maintain habitability on Earth by protecting against deleterious effects of the SW, and reduces atmospheric escape by returning much of the ion outﬂow back to Earth via the Dungey cycle \cite{Dungey1961,Milan2007,Moore2007,Seki2001}. But theoretical models have shown that the magnetic field need not always be protective, and Earth, Mars, and Venus have empirically  similar mass loss rates despite the latter two lacking an intrinsic dipole magnetic field \cite{Blackman2018,Slapak2017,Egan2019,Gunell2018}. When the key question is habitability, planetary evolution must also be considered when assessing the influence of magnetic field \cite{Tarduno2014}. The debate over the protective role of the magnetic field is likely to persist, given the uncertain composition of early planetary atmospheres, solar wind conditions, and the presently limited data from extrasolar examples that bear directly on early solar system conditions.

Here, we develop a dynamical model of the SW-Earth atmosphere interaction to determine whether Earth's atmosphere is a viable extra-solar source of volatiles to the lunar regolith (Figure 2). We first compare the net fraction of Earth's atmospheric and SW flux carried to the Moon by the SW for both unmagnetized (no dynamo) and magnetized (with dynamo) Earth simulation cases. We then combine these with photoionization models for Eoarchean and present-day terrestrial atmospheres, respectively, to infer the relative ratios of these elements at Earth's exobase and predict the abundance ratios in lunar soil. We here address transport to the large regions on the lunar nearside surface without any magnetic anomalies; therefore, mini-magnetospheres of some small, discontinuous lunar magnetized crust are largely irrelevant. The implanted ions may vary spatially, exhibiting higher concentrations at the edges of the anomalies. However, these mini-magnetospheres do not necessarily fully shield the lunar surface from the impinging ions; a higher ram pressure can allow more ions to penetrate, affecting any concentration gradient from the center to the edges \cite{Lianghai2021}. We also assume that the Moon did not have an appreciable global or large-scale magnetosphere during the interval of study. This was assumed in previous modeling \cite{Ozima2005} and has been supported by more recent paleointensity data indicating the absence of any long-lived lunar core dynamo after $\sim 4.36$ Ga \cite{Tarduno2021,Zhou2024}. As we will discuss below, our results can, in turn, be used as a check on this assumption.

\section{Results}

\subsection{Relative total terrestrial atmospheric flux implanted on the Moon for an Eoarchean unmagnetized and contemporary magnetized Earth}

We analyze the planetary and SW components in the outflow using a single-fluid multi-species MHD model dynamically, but with passive advective tracers that separately track Earth's atmosphere material and SW material. The number flux of the terrestrial component is computed using the relation $\mathcal{F}_{\mathsmaller{EW}} = n_{\mathsmaller{EW}}  \left( \vec{v}_{\hspace{-0.2em}\mathsmaller{EW}}\, \cdot \, \widehat{n}_{\mathsmaller{s}} \right)$, while the number flux of the SW component is similarly expressed as $\mathcal{F}_{\mathsmaller{SW}} = n_{\mathsmaller{SW}} \left( \vec{v}_{\hspace{-0.2em}\mathsmaller{SW}}\, \cdot \, \widehat{n}_{\mathsmaller{s}} \right)$. Here $\widehat{n}_{\mathsmaller{s}}$ is the inward lunar surface normal used to calculate the impinging fluxes, $v_{\mathsmaller{SW}}$ and $v_{\mathsmaller{EW}}$ stand for the SW flow speed upstream of the bow shock and the Earth wind (EW) speed, respectively. We refer to the term ``Earth wind'' as the escaping atmospheric ions when the SW interacts with and flows past the terrestrial atmosphere. $n_{\mathsmaller{SW}}$ denotes the number density of the SW, while $n_{\mathsmaller{SW}}$ represents that of EW.

Here (and throughout the text), the term ``magnetized case'' is used for simulation runs where the planet possesses an intrinsic dipolar field with present-day atmosphere. In contrast, the ``unmagnetized case'' signifies the absence of such an intrinsic planetary magnetic field for an early Archean Earth. 

The mass flux received at the Moon over a full lunation ($-\pi$ to $\pi$ rad) exhibits a double-horned structure with global maxima at orbital phase angles $\phi_{\mathsmaller{M}} \sim -\pi/5$ rad and $\sim \pi/5$ rad in the magnetized subcase of Case -- \text{I}, and at phase angles $\phi_{\mathsmaller{M}} \sim -\pi/9$ rad and $\sim \pi/9$ rad in the unmagnetized subcase (Case -- \text{I}). These maxima coincide with lunar crossings of the respective bow shocks (Figures 3 and ED1). We used the phase angle in the Geocentric-Lunar Orbital Plane (GLOP) coordinate frame to indicate the different locations of the tidally-locked Moon around the Earth. The following sign convention is employed for the phase angle: Counterclockwise angles are  positive, while clockwise angles are negative. The phase angle starts at roughly $-\,\pi$ rad near new Moon when it is in the dayside upstream SW and progresses to $0$ rad at full Moon in the Earth's magnetotail. The Moon then moves from $0$ rad to near $+\,\pi$ rad during the remainder of the cycle (Figure 2).

The primordial Sun (prior to $\sim 4.2$ Ga) likely emitted a stronger SW \cite{Shaviv2003}. Only at such early times could the Earth have possibly been unmagnetized \cite{Tarduno2023}. We therefore vary the SW parameters between the magnetized and unmagnetized cases, where the latter could represent ancient SW conditions (Tables ED1 and ED2). However, in Case -- \text{III} (Table 1), representing a period when the existence of a dynamo is uncertain, we use identical SW pressure for both the magnetized and unmagnetized subcases to isolate the influence of the field on atmospheric escape. The surface density of the terrestrial atmosphere is not fixed between the magnetized and unmagnetized cases, but emerges from assuming the same total atmosphere mass for the two cases.

The fluxes are computed at the  nearside sub-Earth point on the Moon. The Apollo soil data to which we compare our results are also from that vicinity \cite{Heiken1991,Orloff2006}. For over half of its orbit, the Moon is in the SW plasma on the sunlit side of its orbit, and outside of the bow shock. For both the magnetized and unmagnetized cases, we therefore observe that the SW flux increases slowly when the Moon is at  first quarter and falls to zero after the third quarter when the sub-Earth point is on the night side. As the Moon ingresses or egresses the magnetosphere, the SW number density abruptly increases at the bow shock, causing in the double-peaked structure. Within the magnetotail, the SW flux drops by an order of magnitude. On the other hand, the EW, originating from the terrestrial atmosphere, only adds to the flux within the magnetotail region.

The orbit-averaged species-total SW and terrestrial atmosphere fluxes impinging upon the lunar surface are tabulated in Table 1. In each case, the total atmospheric mass is kept the same to isolate the comparison of SW interaction with a magnetic versus non-magnetic Earth. The Moon is believed to have tidally locked to Earth shortly after its formation, and  subsequently slowly drifted away from Earth at $\sim 15\, R_{\mathsmaller{\oplus}}$  to the current distance of $\sim 60\, R_{\mathsmaller{\oplus}}$ \cite{Farhat2022}. We used the current Earth-Moon separation of $60\, R_{\mathsmaller{\oplus}}$ to calculate SW and EW fluxes for the magnetized cases, denoting the present time and a lunar distance of $40\, R_{\mathsmaller{\oplus}}$ for the unmagnetized cases. This lunar distance was attained around 4.0 Ga, which falls within the constraints on the earliest presence of the geodynamo of $\sim 3.5$ Ga based on paleomagnetic studies of extant rocks \cite{Tarduno2010,Biggin2011,Usui2009}, and approximately $\sim 4.2$ Ga based on paleointensity data from detrital zircons of Australia and South Africa \cite{Tarduno2015,Tarduno2020,Bono2022,Tarduno2023}.

For the fiducial case (Case -- \text{I}), we utilized the present SW velocity and density values for the magnetized sub-case, and the current terrestrial atmosphere profile of the contemporary Earth. For the unmagnetized sub-case, we used the ancient SW parameters, representing the ancient Sun-Earth-Moon system, and determined the surface density of the atmosphere by conserving the total atmospheric mass from the magnetized case.

The Earth's atmosphere contribution averaged over one lunar orbit (referred to as ``planetary flux'') for the current Earth, with an intrinsic dipole field, is comparable to  that observed when there is no intrinsic magnetic field. This results because the magnetic field distends the same amount of atmosphere mass onto a power-law tail rather than an exponential one, allowing more of Earth's atmosphere to survive at larger radii from stripping, which compensates for any protection the field might provide. This adds to the reasons why a magnetic field need not always be protective \cite{Blackman2018,Egan2019}. In Case -- \text{III}, however, the ram pressure of the SW, albeit stronger than in Case -- \text{I}, is kept constant by fixing the values of SW number density and velocity.  In Case -- \text{III}, we isolated the magnetic field effect. The EW flux is an order of magnitude smaller in the magnetized subcase, suggesting an overall modest net protection effect. 

In Case -- \text{IV}, the atmosphere surface density for a magnetized Earth is reduced by five orders of magnitude while maintaining the same SW values compared to Case -- \text{I}. Since this reduction decreases the atmospheric mass available to be picked up by the SW, we observe a corresponding decrease in the escaping EW flux. However, the decrease in flux is not proportional to the reduction in base density. This is because the atmosphere's surface density does not scale linearly with the outflowing planetary flux, as further discussed in the next section. 

By further reducing the surface densities in both magnetized and unmagnetized scenarios in Case -- \text{II}, and increasing the ram pressure of the incoming wind, we observe that the planetary flux is $2.41 \times 10^{8}\; \mathrm{m^{-2}\, s^{-1}}$ for a magnetic Earth and $3.10 \times 10^{9}\; \mathrm{m^{-2}\, s^{-1}}$ for a non-magnetic Earth, which are higher than any of the previously discussed cases, suggesting that the SW ram pressure plays a more significant role in enhancing the value of the EW flux than the terrestrial mass density in the interaction region close to the respective plasma or magnetopauses. This is more clearly elucidated in Case -- \text{V}, where the Earth's atmosphere surface density is increased back to its fiducial  value, along with an increase in both SW density and velocity, resulting in a higher average EW flux. 

Overall, based on simulation data in Table 1, we observe that the relative fraction of EW flux compared to the SW flux is not strongly dependent on the magnetic field for most cases except for Case -- \text{III}. In that case, we have isolated the effect of the magnetic field change only, and see an order of magnitude reduction in planet flux (column 6).  

\subsection{Specific Ion Species implanted from Earth Atmosphere}

We estimate the contribution of N and light noble gas fluxes to the lunar soil from both SW and EW using the photoionization model 
(Tables ED3–ED5). Comparing curves for various ions for the magnetized and unmagnetized cases (Figures 4 and ED2), we infer that the unmagnetized phases do not make a substantial additional contribution to EW deposition compared to the magnetized ones. The various cases for magnetic and non-magnetic Earths are described in Table 1.

The SW effectively captures and mixes ions in the EW-SW interaction region without fractionation by mass, because the ion gyro-radii are small compared to global scales. The pick-up EW ions thereby undergo rapid acceleration, all reaching the same average velocity as the SW. Some of these accelerated ions implant in the lunar soil. Because all enter at the same velocity as the SW and the plasma is collisionless before impact, heavier ions from both EW and SW origins may penetrate more deeply than lighter ions, but EW and SW components would not separately be segregated by depth alone. The depth profile may, however, reflect the evolution of Earth's atmosphere on geological timescales \cite{Hashizume2000,Tucker2019}.

\subsection{Implications of the variation in isotopic composition of Lunar soil due to implanted solar and Earth winds}

In order to determine whether the non-solar components of the lunar soil have a terrestrial origin attributable to the EW, we compare the theoretically calculated implanted EW with the observed non-solar fluxes for the present-day magnetized Earth with those of non-magnetized early Earth. Our MHD simulations for the magnetized and unmagnetized phases of Earth suggest that mixing between two primary end members, solar and planetary, controls the lunar geochemical variation (Figures 5 and ED3). The non-solar contribution to the lunar soil is then estimated and compared to the EW flux computed using our theoretical model, as tabulated in Table 2. Theoretical mixing diagrams (Figures 6 and ED4) characterize the relative proportions of various isotopes sourced from SW and a non-solar component (NSC), as well as the elemental abundances of different species. The construction and interpretation of these diagrams are explained in detail in the \nameref{sec:Methods} section.

The respective color bars in the mixing plots (Figures 6 and ED4) show increments of the SW mass fraction in the mixture. Stated differently, they represent the weight fraction due to the contributions of the two elements from the two end members. Comparing the corresponding subplots in Figures 6 and ED4 suggests that the SW mass fraction decreases from unity more rapidly for an unmagnetized Earth with an early Archean atmosphere (and less steeply for the contemporary terrestrial atmosphere) as one traverses away from the SW component point, indicated by the orange square. This, in turn, implies that in the present-day atmosphere, there is a significant amount of mixing between the SW and the non-solar component. Conversely, the mixture in the early Archean Earth case is predominantly solar. Using the computed non-solar fraction for each species, we estimated the non-solar flux by multiplying it with the SW flux values. This estimated non-solar flux—which is a measure of the observed species abundances from the Apollo sample data—was then compared with the EW flux computed using the combined MHD-ionization model for both the magnetized and unmagnetized Earth phases. The combined model explains the data only for the contemporary magnetized Earth case when the EW flux value surpasses the recorded non-solar flux (Table 2).

We utilized isotope ratio data of N and H in lunar regolith grains from Hashizume et al. \cite{Hashizume2000} to compare with our theoretical mixing curves between the two fixed end-members, SW and EW, for different exobase heights in both magnetized and unmagnetized cases. Hashizume et al. \cite{Hashizume2000} attributed the isotopic variation in lunar nitrogen, and the depletion of $\mathrm{{}^{15}N}$ relative to terrestrial nitrogen, to the mixing of a solar and a non-solar component. They assigned the latter to presolar interstellar solids that underwent $\mathrm{{}^{15}N}$ enrichment through isotopic fractionation. However, our combined 3D-MHD and ionization model predicts that this implanted non-solar N could have originated from the terrestrial atmosphere if the exobase altitude was below $300$ km. This is evident from the mixing plot, where the majority of lunar N and H isotope sample data lie close to the curves for exobase heights less than $300$ km (Subfigures 6 (a) \& ED4 (a)). 

The data for other light volatile elements, such as He, Ne, and Ar in lunar ilmenites and regolith breccias, were obtained from Heber et al. \cite{Heber2003}, who focused on the temporal variation of these isotopic compositions and revealed a strong correlation among these ratios. Given the absence of known processes capable of temporally altering Ne isotopic composition in the SW, they argued that the variation in the He isotopic ratio cannot be explained by incomplete H burning and mixing in the Sun’s outer convective zone, which closely aligns with the protosolar composition. Heber et al. \cite{Heber2003} thus attributed these variations to secondary processes such as grain surface erosion and diffusion. Once again, our theoretical calculations show that these implanted light element gases can also be well explained by the mixing between the SW and EW. Our model best matches the experimental data from lunar soil if the exobase height is $\sim 250$ km in Case -- \text{I} for a magnetic Earth (refer to Subfigures 6 (a) \& 6 (b)) and $\sim 275$ km in Case -- \text{I} for a non-magnetic Earth (see  Subfigures ED4 (a) \& ED4 (b)). 

Similarly, the observed correlation between He and Ar isotopic ratios \cite{Heber2003} can also be accounted for by both our present-day magnetized (Subfigure 6 (c)) and Eoarchean unmagnetized (subfigure ED4 (c)) atmosphere models that include part SW and part EW transported constituents if the exobase height is close to $190$ km. The lunar data also match our binary mixing curves for N-H isotope mixing, exhibiting greater curvature compared to that for He-Ne and He-Ar isotope mixing, as indicated by the mixing diagrams.

From columns 4 and 5 of Table 2, we infer that all species, except H, can be accounted for by terrestrial components transported in the form of EW in the magnetized present-day case but not in the Eoarchean case (Case -- \text{I}). The Eoarchean model predicts too dominant a SW contribution to species-specific abundance compared to the data due to the intense SW that increases the ratio of SW to EW components around 3.8 Ga ago (refer to columns 8 and 9 of Table 2). That neither model matches the observed lunar H implies that the observed lunar H is non-terrestrial and could very well have a solar or extrasolar origin. Our model's mixing ratio curves (Figures 6 and ED4) are strongly sensitive to the exobase height. Although this height is difficult to measure empirically, the sensitivity of the mixing ratio observables to it serves as a theoretical constraint on its evolution. 

\section{Discussion}

Our findings reveal that Earth's atmosphere contributes significantly to light volatile elements on the lunar regolith when the Moon is situated in the magnetotail. The calculations, using a combination of 3-D MHD numerical simulations of magnetized SW-Earth atmosphere interactions and a semi-analytic ionization model, demonstrate that material from a present day Earth's atmosphere can indeed account for the non-solar contribution to the N and noble gases constituents in lunar soil, offering a plausible explanation for the observed isotope ratio differences between solar and non-solar sources. 

Only a primitive Earth phase could be unmagnetized, a time when the SW ram pressure was much stronger. But we find that the relative SW and EW abundances fraction are under predicted by such an early non-magnetic Earth primarily due to increased SW power at early times. Our model predictions for the lunar soil therefore suggest that its composition reflects a dominant acquisition during the active dynamo phase of Earth. This is, in turn, consistent with the flatter temporal evolution profiles of the SW mass loss rate \cite{Vidotto2021}, where the time-integrated mass deposition is governed by present-day conditions.

Ozima et al. \cite{Ozima2005} assigned an age of 3.8 - 3.9 Ga to the ilmenites used for volatile measurements. This is a common age for some high-Ti lunar basalts that might be the original hosts of the ilmenites and overlaps with some estimates of the late heavy bombardment \cite{Niem2012}, which could disrupt the hosts and place the ilmenites on the surface for ion implantation. Our results indicate that if the implantation ages are this old, the Moon did not have a core dynamo or magnetosphere, as this would block terrestrial atmospheric transfer needed to exemplify the ilmenite volatile data (Figures 5, 6, ED3, and ED4). Nevertheless, our model is also consistent with much later implantation ages.

We have also isolated the effect of a magnetic field to assess its influence when the presence or absence of a dynamo might be unknown. Keeping a fixed total atmosphere mass and fixed solar wind properties, we found that the magnetic field reduces the escaping EW fraction by about an order of magnitude. This effect underscores the subtle interplay between the SW, Earth's dynamo, and the Moon's orbital characteristics when assessing the net influence of the magnetic field.

We conclude that the delivery of EW ions, sourced from the terrestrial atmosphere, to the Moon occurred during Earth's extended magnetized phase, as opposed to any brief, putative non-magnetic phase during early Earth. Our results show that the observed soil isotope mixing ratio data lie on hyperbolae of fixed exobase altitude, with slightly different values for different element ratios, but with widely varying total EW to SW fractions. This suggests that the latter varied over Earth's history more than the former. Comparing the model to the data suggests that the lunar soil never acquired substantial material when the exobase was less than $190$ km.
\clearpage

\section{Methods}
\label{sec:Methods}

\subsection{Wind-Atmosphere MHD Interaction Model and Computational Methodology}

We model the intrinsic magnetic field  of Earth by the far-field solution of an Earth-centered point-like dipole
with  dipole moment $8.07 \times 10^{15}$ $\mathrm{T-m^3}$ \commentout{$8.068 \times 10^{15}$ $\mathrm{T\,m^3}$}. The confined magnetic vector potential $\vec{A}_{\mathsmaller{GF}}$ (where $\vec{B}_{\mathsmaller{GF}} = \vec{\nabla} \times \vec{A}_{\mathsmaller{GF}}$) in the initial ($t = 0$) frame is defined as
\begin{equation}
  \vec{A}_{\mathsmaller{GF}} = \left \{
  \begin{aligned}
    &8\,B_{\mathsmaller{\oplus}}\;r\;\sin{\theta}\;\left[1-\frac{r}{\widetilde{R}}\right]^{2}\;\widehat{\phi}, && \text{if}\ r \le \frac{R_{\mathsmaller{\oplus}}}{2} \\[10pt]
    &B_{\mathsmaller{\oplus}}\;r\;\sin{\theta}\;\left(\frac{R_{\mathsmaller{\oplus}}}{r}\right)^{3}\;\left[1-\frac{r}{\widetilde{R}}\right]^{2}\;\widehat{\phi}, && \text{if}\ r > \frac{R_{\mathsmaller{\oplus}}}{2}
  \end{aligned} \right.
  \label{eq:eq1}
\end{equation}
where $\widetilde{R} = 20\,R_{\mathsmaller{\oplus}}$ is the initial B-field cutoff boundary, $R_{\mathsmaller{\oplus}} = 6.371$ Mm is  Earth's radius, and $B_{\mathsmaller{\oplus}}$ is the mean value of the magnetic field at the magnetic equator on the Earth's surface. Since the SW is magnetized, the outside ambient (interplanetary) field completely drapes around the magnetopause, and the magnetosphere is surrounded by the ambient wind magnetic field. As the simulation progresses, the field lines are compressed on the dayside by both the SW dynamic and thermal pressures; on the nightside it extends into an open magnetotail-like configuration (Figure 1).

\subsubsection{Isothermal Atmosphere Model}
In the frame rotating with a constant angular velocity, $\vec{\Omega}$, the fully compressible inviscid momentum equation (Euler form) can be written
\begin{equation}
\begin{split}
    \frac{\partial \vec{v}}{\partial t} + (\vec{v} \cdot \vec{\nabla})\,\vec{v} & =  \frac{\partial \vec{v}}{\partial t} + \vec{\nabla}\left(\frac{v^2}{2}\right)\,-\,(\vec{v} \times \vec{\nabla} \times \vec{v}) \\
   & = -2\Omega \times \vec{v} - \vec{\Omega} \times \vec{\Omega} \times \vec{r} -\frac{\vec{\nabla}p}{\rho} + \vec{g} \\
\end{split}
\label{eq:eq2}
\end{equation}
Here $\vec{v}$ is the velocity in the rotating frame, $\vec{r}$ is the radius vector, $p$ is the pressure, $\rho$ is the mass density of the atmosphere, and $\vec{g}$ is the surface gravity. The quantities appearing after the second equality are source and sink terms that provide the acceleration. A useful solution of \autoref{eq:eq2} can be obtained by approximating the atmosphere to be in hydrostatic equilibrium,  which is valid for highly subsonic atmospheric flow. Then Earth's gravity and  atmospheric pressure gradient are the dominant forces in balance. For an isothermal atmosphere obeying the ideal gas law, with gravity acting radially inward, $\vec{g} = -g\hat{r}$, and no angular dependence of density or pressure, we have
\begin{equation}
    \vec{\nabla}p = -\rho g\hat{r} \Longrightarrow \frac{p}{p_{\circ}} = \frac{\rho}{\rho_{\circ}} = e^{-\frac{r-R_{\mathsmaller{\oplus}}}{H}}
    \label{eq:eq3}
\end{equation}
where $p_{\circ}$ and $\rho_{\circ}$ are the surface pressure and density respectively, and $H$ is the scale height.

The Earth's magnetic field of order 50 $\muup$T near the surface produces a magnetic pressure $p^{\mathsmaller{(Mag)}}$ of $\sim 10^{-3}$ Pa ($\simeq 10^{-8}$ atm). At the outer reaches of the dayside magnetosphere where the magnetic pressure starts to dominates Earth's atmospheric pressure, the nearly vacuum dipole magnetic field determines the global plasma structure, and the characteristic exponential fall-off of the isothermal density profile is no longer valid. In this low plasma-beta ($\upbeta_{\mathsmaller{p}}$) regime, where $\upbeta_{\mathsmaller{p}}$ is the ratio of thermal to magnetic pressure, we model the atmospheric density as a power law $\rho\propto r^{-\xi}$ with index $\xi > 0$. Specifically, equating the two pressures we obtain the radius $(r_{\mathsmaller{\beta_p}})$ where $\upbeta_{\mathsmaller{p}} = 1$,
\begin{equation}
   p^{\mathsmaller{(Atm)}} = p^{\mathsmaller{(Mag)}} \Longrightarrow  p_{\circ}\; e^{-\frac{r_{\mathsmaller{\beta_p}}-\;R_{\mathsmaller{\oplus}}}{H}} = \frac{p_{\mathsmaller{1}}}{\left[\frac{r}{r_{\mathsmaller{\beta_p}}}\right]^{\xi}} \approx \frac{B_{\mathsmaller{\oplus}}^2}{2\,\mu_{\circ}} \left(\frac{R_{\mathsmaller{\oplus}}}{r}\right)^6
    \label{eq:eq4}
\end{equation}
$p_{\mathsmaller{1}}$ represents the atmosphere thermal pressure at that transition radius. We use the following pressure profile for the atmosphere in the magnetized case:
\begin{equation}
  p_{\mathsmaller{M}}^{\mathsmaller{(Atm)}} = \left \{
  \begin{aligned}
    &p_{\circ,{\mathsmaller{M}}}\; e^{-\frac{r-\;R_{\mathsmaller{\oplus}}}{H_{\mathsmaller{M}}}}, && \text{if}\ R_{\mathsmaller{\oplus}} <\,r \le\, r_{\mathsmaller{\beta_p}} \\[10pt]
    &\frac{p_{\mathsmaller{1},{\mathsmaller{M}}}}{\left[\frac{r}{r_{\mathsmaller{\beta_p}}}\right]^{\xi}}, && \text{if}\ r_{\mathsmaller{\beta_p}} <\,r \le   r_{\mathsmaller{MP}} \\[10pt]
  \end{aligned} \right.
  \label{eq:eqAtmMag}
\end{equation}
Here $H_{\mathsmaller{M}}$ is the atmospheric scale height during the current magnetized Earth phase. In the case with no planetary dipole field, atmospheric pressure is assumed to vary as
\begin{equation}
  p_{\mathsmaller{UM}}^{\mathsmaller{(Atm)}} = p_{\circ,{\mathsmaller{UM}}}\; e^{-\frac{r-\;R_{\mathsmaller{\oplus}}}{H_{\mathsmaller{UM}}}},\hspace{0.1in} \text{for}\ R_{\mathsmaller{\oplus}} <\,r \le\, r_{\mathsmaller{PP}} 
  \label{eq:eqAtmUnmag}
\end{equation}
$H_{\mathsmaller{UM}}$ is the scale height of the Archean atmosphere during the unmagnetized era. The plasmapause boundary ($r_{\mathsmaller{PP}}$) and the pressure at the base of the planet's atmosphere ($p_{\circ,{\mathsmaller{UM}}}$) in the unmagnetized case is determined by equating the total atmospheric mass to that in the magnetized case.
\begin{equation}
  \EuScript{M}_{\mathsmaller{M}} = \rho_{\circ,{\mathsmaller{M}}} \int\limits_{R_{\mathsmaller{\oplus}}}^{r_{\mathsmaller{\beta_p}}} r^2\, e^{-\frac{r-\;R_{\mathsmaller{\oplus}}}{H_{\mathsmaller{M}}}}\, dr\, + \, \rho_{\mathsmaller{1},{\mathsmaller{M}}} \int\limits_{r_{\mathsmaller{\beta_p}}}^{r_{\mathsmaller{MP}}} r^{2-\xi}\;r_{\mathsmaller{\beta_p}}^{\xi} \, dr \, = \, \rho_{\circ,{\mathsmaller{UM}}} \int\limits_{R_{\mathsmaller{\oplus}}}^{r_{\mathsmaller{PP}}} r^2\, e^{-\frac{r-\;R_{\mathsmaller{\oplus}}}{H_{\mathsmaller{UM}}}}\, dr = \EuScript{M}_{\mathsmaller{UM}}
  \label{eq:masscons}
\end{equation}
Balancing the atmospheric thermal pressure with the SW ram and thermal pressures yields
\begin{equation}
  r_{\mathsmaller{PP}} = H_{\mathsmaller{UM}}\; \log_{\mathsmaller{e}}\left(\frac{p_{\circ,{\mathsmaller{UM}}}}{\rho_{\mathsmaller{SW}}\,v_{\mathsmaller{SW}}^2\, + \,  n_{\mathsmaller{SW}}\,k_{\mathsmaller{B}}\,T_{\mathsmaller{SW}}}\right)
  \label{eq:ppbound}
\end{equation}
We integrate the last term in \autoref{eq:masscons} analytically. Simplifying the expression results in
\begin{equation}
  \Upxi \, -\, \varsigma_{\mathsmaller{PP}} \,= \, \log_{\mathsmaller{e}}\upvarpi
  \label{eq:transcend}
\end{equation}
where:
\begin{align*}
 \Upxi & =  \log_{\mathsmaller{e}}\left(\frac{\EuScript{M}_{\mathsmaller{M}} \, k_{\mathsmaller{B}}\,T_{\circ}}{\mu_{\mathsmaller{A}}\;m_{\mathsmaller{H}}\,R_{\mathsmaller{\oplus}}^2\, H_{\mathsmaller{UM}}} \times \frac{1}{\rho_{\mathsmaller{SW}}\,v_{\mathsmaller{SW}}^2\, + \,  n_{\mathsmaller{SW}}\,k_{\mathsmaller{B}}\,T_{\mathsmaller{SW}}}\right)    \\[0.8em]
 \varsigma_{\mathsmaller{PP}} & = \frac{r_{\mathsmaller{PP}}}{H_{\mathsmaller{UM}}}    \\[0.8em]
 \upvarpi & = \left(\frac{1}{\varsigma_{\mathsmaller{PP}}}\right)^2 \left[2 - e^{-\varsigma_{\mathsmaller{PP}}} \left(\varsigma_{\mathsmaller{PP}}^2 + 2\,\varsigma_{\mathsmaller{PP}} + 2\right)\right] \, + \, \frac{2}{\varsigma_{\mathsmaller{PP}}} \left[1 - e^{-\varsigma_{\mathsmaller{PP}}} \left(1 + \varsigma_{\mathsmaller{PP}}\right)\right] \, + \, \sinh{\varsigma_{\mathsmaller{PP}}}\, -\, \cosh{\varsigma_{\mathsmaller{PP}}} + 1
\end{align*}
$\mu_{\mathsmaller{A}}$ denotes the mean molecular mass and $m_{\mathsmaller{H}}$ is the hydrogen mass. The transcendental equation above is solved numerically to obtain $r_{\mathsmaller{PP}}$, $p_{\circ,{\mathsmaller{UM}}}$, and $\rho_{\circ,{\mathsmaller{UM}}}$.

\subsubsection{Parametric analysis of the magnetosphere standoff distance}
The magnetosphere standoff distance can be computed from the pressure balance equation by equating the ram pressure of the SW
plus its magnetic and thermal pressures, to the thermal pressure of the Earth's atmosphere and the magnetic pressure of the dipolar geomagnetic field. Explicitly, this gives
\begin{equation}
   \frac{\rho_{\mathsmaller{SW}}\,v_{\mathsmaller{SW}}^2}{2}\,+ \, \frac{B_{\mathsmaller{SW}}^2}{2\,\mu_{\circ}} \,+ \,\frac{\rho_{\mathsmaller{SW}}\,v_{\mathsmaller{Th,\,SW}}^2}{2}\, =  \frac{p_{\mathsmaller{1}}}{\left[\frac{r}{r_{\mathsmaller{\beta_p}}}\right]^{\xi}} \,+ \,  \kappa\frac{B_{\mathsmaller{GF}}^2}{2\,\mu_{\circ}}
    \label{eq:eq6}
\end{equation}
where $\rho_{\mathsmaller{SW}} \approx n_{\mathsmaller{SW}}\; m_{\mathsmaller{P}}$ is the mass density of the SW, $v_{\mathsmaller{Th,\,SW}}$ is the SW thermal velocity, $B_{\mathsmaller{SW}}$ is the magnetic field strength of the oncoming wind, and $B_{\mathsmaller{GF}}$ (\autoref{eq:eq1}) denotes the geomagnetic field strength.  Here $r_{\mathsmaller{MP}}$ stands for the location of the terrestrial magnetopause, and $\kappa \approx 2$ is the dipolar magnetic field compression factor \cite{Gombosi2004}.

\subsubsection{Solar and planetary parameters}
Table ED1 lists the basic model parameters used in the simulations for the contemporary Sun and Earth. The parameters characterizing the stellar and planetary outflows are the escape velocity parameter and the plasma-beta in the SW-atmosphere interaction region. The atmospheric escape speed is given by $\left[ 2\,G\,M_{\mathsmaller{\oplus}} / R_{\mathsmaller{\oplus}}\right]^{\frac{1}{2}}$, while the sound speed at the base of the planet atmosphere is $c_{\mathsmaller{s}}^{\mathsmaller{(Atm)}} = \left[ \gamma \, p^{\mathsmaller{(Atm)}}(r) / \rho^{\mathsmaller{(Atm)}}(r) \right]^{\frac{1}{2}} = \left[ \gamma\,k_B\,T_{\circ} / m_{\mathsmaller{mol}}^{\mathsmaller{(Atm)}}(r)\right]^{\frac{1}{2}}$. The escape velocity parameter is then defined as 
\begin{equation}
\Zhe^{\mathsmaller{(Atm)}} = \left[v_{\mathsmaller{Esc}}^{\mathsmaller{(Atm)}} /  c_{\mathsmaller{s}}^{\mathsmaller{(Atm)}}\right]^{2} = \text{Ma}_{\mathsmaller{\text{p}}}^2
\end{equation}
where $\text{Ma}_{\mathsmaller{\text{p}}}$ is the plasma Mach number. The plasma-beta characterizes the dynamical significance of the magnetic field and is given by 
\begin{equation}
\upbeta_{\mathsmaller{\text{p}}}^{\mathsmaller{(Atm)}} =2\,\mu_{\circ}\, p^{\mathsmaller{(Atm)}}(r) / B_{\mathsmaller{GF}}^2(r)
\end{equation}
Similarly, corresponding expressions for $v_{\mathsmaller{Esc}}^{\mathsmaller{(SW)}},\,c_{\mathsmaller{s}}^{\mathsmaller{(SW)}},\, \Zhe^{\mathsmaller{(SW)}}$, and $\upbeta_{\mathsmaller{\text{p}}}^{\mathsmaller{(SW)}}$ are used for the SW.

\subsubsection{Planet interior model}
The planet surface is treated as an internal boundary and the planet's interior is not dynamically evolved during the simulation runs. The interior parameter values are kept fixed by overwriting them at each time step. Since the interior does not interact with the SW, its physical conditions are computationally unimportant and we model it as an isothermal sphere. The gravitational acceleration due to the planetary core is then given by
\begin{eqnarray}
\vec{g}(x, y, z) = -\frac{4}{3}\pi\, G\, \rho_{\mathsmaller{\oplus}}\, r\hat{r} = -\frac{4}{3}\pi\, G\, \rho_{\mathsmaller{\oplus}}\,\left[ x\hat{x}  +  y\hat{y}  +  z\hat{z}  \right]
\label{eq:eq7}
\end{eqnarray}
Assuming hydrostatic equilibrium, the pressure $p_{\mathsmaller{\oplus}}$ inside the spherical volume, caused by gravitational compression, as a function of the radial distance $r$ is
\begin{eqnarray}
\frac{1}{p_{\mathsmaller{\oplus}}(r)} = \frac{1}{p_{\circ}} + \frac{2}{3}\pi\, G\, \left(\frac{\mu_{\mathsmaller{A}}\;m_{\mathsmaller{H}}}{k_{\mathsmaller{B}}\,T_{\circ}}\right)^2\, \left[r^2 - R_{\mathsmaller{\oplus}}^2 \right]
\label{eq:eq8}
\end{eqnarray}
where $p_{\circ}$ is the atmosphere surface pressure. Finally, for our magnetized planet runs, we approximate the Earth as a roughly uniformly magnetized sphere on the interior $\left(r \le R_{\mathsmaller{\oplus}} / 2 \right)$ with a smooth magnetic field close to the center, where singularity occurs and retain the dipolar structure outside $r > R_{\mathsmaller{\oplus}} / 2 $ (see \autoref{eq:eq2}). At large distances where the planet's intrinsic dipole field would be very weak, we impose a cutoff boundary ($\widetilde{R} = 20\, R_{\mathsmaller{\oplus}}$) outside of which the field vanishes. 

\subsubsection{On the MHD approximation and  requisite resolution}
Even though the Earth and SW plasmas  are collisionless, the gyro-radii of both, using  SW field strengths, are smaller than the macroscopic scales of interest, such as the distance from Earth to the bow shock. This crudely justifies simulating the interaction between the SW and atmosphere using the MHD approximation to gain some basic insights.

The numerical simulations must resolve a pressure scale height of the planetary atmosphere that is small compared to dynamical scales of interest, but not too small to present practical computational limitations. The pressure scale heights  equal the density scale heights for the isothermal equations of state that we employ here. Since 3-D numerical MHD simulations are computationally expensive, we model the star-planet interaction using  a grid  with multiple levels of refinement and finely resolve only the regions of interest. We set up a static, locally-refined grid in a planetocentric frame of reference so that the location of the planet remains fixed throughout the run. Working in the planet frame also simplifies the treatment of the planet's core as a fixed bounded region that does not evolve in the simulations. 

We carry out our global MHD simulations with \href{https://bluehound2.circ.rochester.edu/astrobear/}{\normalsize{AstroBEAR}} \cite{Cunningham2009,Jonathan2013}, a massively parallelized adaptive mesh refinement (AMR) code that includes a variety of multiphysics solvers, such as self-gravity, magnetic resistivity, heat conduction, radiative transport, and ionization dynamics\footnote[2]{\href{https://bluehound2.circ.rochester.edu/astrobear/}{http://astrobear.pas.rochester.edu/}}. AstroBEAR has been previously utilized, benchmarked, and tested for modeling the interaction between planetary atmospheres and stellar winds \cite{Shule2013,Nellenback2016,Atma2021}. We solve the MHD equations in the planetocentric reference frame on a Cartesian grid in the following conservative  form:
\begin{eqnarray}
\frac{\partial \rho }{\partial t} + \vec{\nabla } \cdot \left[\rho \vec{v}\right] = 0
\label{eq:eq10}
\end{eqnarray}
\begin{eqnarray}
\frac{\partial}{\partial t} \left(\rho \vec{v}\right) + \vec{\nabla } \cdot \left[\rho (\vec{v} \otimes \vec{v}) -  \vec{B} \otimes \vec{B} + \left(p + \frac{B^2}{2}\right) \stackrel{\leftrightarrow}{I} \right]= \rho \vec{g}
\label{eq:eq11}
\end{eqnarray}
\begin{equation}
\frac{\partial E}{\partial t} + \vec{\nabla } \cdot \left[ \left(E + p + \frac{B^2}{2}\right) \vec{v}\: - \:\vec{B} \left(\vec{B} \cdot \vec{v} \right) \right] = \rho \vec{v} \cdot \vec{g}
\label{eq:eq12}
\end{equation}
and
\begin{equation}
\frac{\partial \vec{B}}{\partial t} - \vec{\nabla } \times \left(\vec{v} \times \vec{B}\right) - \cancelto{0}{\frac{1}{\mu_o\, \sigma}\nabla^2 \vec{B}} = 0
\label{eq:eq13}
\end{equation}
where $\rho$ is the mass density, $\vec{v}$ is the fluid velocity, and the magnetic field is given by $\vec{B}$. Equations \ref{eq:eq10} to \ref{eq:eq13} are written in rationalized electromagnetic units (in which the magnetic permeability $\mu =1$) as defined by Cunningham et al. \cite{Cunningham2009}. \autoref{eq:eq10} is the continuity equation, which describes the conservation of plasma mass. The conservation of momentum density $(\rho \vec{v})$ in the fluid is given by \autoref{eq:eq11}. The quantity $P_{\mathsmaller{T}} = p + \frac{B^2}{2}$ denotes the total pressure, consisting of the thermal pressure of the gas $(p)$ and the magnetic pressure. $\vec{g}$ is the gravitational acceleration experienced by the fluid due to the field of the planet. \autoref{eq:eq12} represents the evolution of the total energy density $E$ $\left(E = \frac{p}{\gamma - 1} + \frac{1}{2} \rho v^2 + \frac{B^2}{2}\right)$, which is the sum of internal, kinetic, and magnetic energy densities, respectively. The  magnetic induction associated with fluid motion is described by \autoref{eq:eq13}. 

We solve  these ``ideal'' MHD equations, i.e. without explicit viscous or resistive terms to maximize the dynamic range because the phenomenon we are interested in, namely, the atmospheric escape due to the electromagnetic star-planet interaction  are at least in principle not much affected by magnetic field diffusion into the upper atmosphere \cite{Russell2006}. Throughout most of the magnetosphere, collision frequencies are sufficiently low to be negligible when compared to the 
ion gyro-frequency, and an estimate for transport is the Bohm diffusion coefficient ($D_{\mathsmaller{B}}$), which has the form
\begin{equation}
   D_{\mathsmaller{B}} \sim n\, T_{\mathsmaller{i}}^{-\frac{3}{2}} \, r_{\mathsmaller{g},\,\mathsmaller{i}}^{\;\; 2} \, \gg \nu_{\mathsmaller{m}} = \dfrac{\eta}{\mu_{\circ}}
    \label{eq:eq36}
\end{equation}
where $r_{\mathsmaller{g},\,\mathsmaller{i}}$ is the ion gyroradius, $T_{\mathsmaller{i}}$ denotes the ion temperature of the plasma, and $\nu_{\mathsmaller{m}}$ is the classical magnetic diffusivity. This (Bohm) is the fastest possible diffusion, applicable only when the mean free path approaches the ion gyroradius. However, the magnetospheric plasma is essentially collisionless, with the ion mean free path being much larger than the typical size of the magnetosphere. The magnetic Reynolds number (the ratio between the advective and diffusive terms of the induction equation) provides an estimate of the comparative effects between the induction of the magnetic field by the motion of the conducting plasma and magnetic diffusion:
\begin{equation}
\begin{split}
    Re_{\mathsmaller{m}} & = \dfrac{\abs{\vec{\nabla } \times (\vec{v} \times \vec{B})}}{\abs{\dfrac{\eta\,\nabla^2 \vec{B}}{\mu_{\circ}}}} \sim \dfrac{\mu_{\circ} \, v_{\mathsmaller{f}} \, L_{\mathsmaller{S}}}{\eta}  \sim \dfrac{B\, L_{\mathsmaller{B}}}{\eta} \sqrt{\dfrac{\mu_{\circ}}{\rho}} \\[2ex]
    & \sim \dfrac{6.71 \times 10^{12}}{\mathrm{ln}\,\Lambda}\, \left(\dfrac{n}{1\, \mathrm{m^{-3}}}\right)^{-\frac{1}{2}} \left(\dfrac{B}{1\, \mathrm{nT}}\right)\,\left(\dfrac{L_{\mathsmaller{S}}}{1\, \mathrm{Mm}}\right)\,\left(\dfrac{T_{\mathsmaller{i}}}{10^{5}\, \mathrm{K}}\right)^{\frac{3}{2}} 
    \label{eq:eq35}
    \end{split}
\end{equation}
Here, $\mathrm{ln}\,\Lambda$ is the Coulomb logarithm which characterizes the effectiveness of Coulomb collisions between charged particles in the weakly coupled plasma, $\eta$ is the Spitzer resistivity, and $L_{\mathsmaller{S}}$ denotes the typical length scale of the magnetospheric flow. In the second term on the right, we assumed the outflow velocity ($v_{\mathsmaller{f}}$) to be the characteristic velocity scale for fully developed magnetohydrodynamic motion in the system. Typically, $Re_{\mathsmaller{m}} \sim 10^{15}$ (corresponding to $L_{\mathsmaller{S}} \sim 10\,\mathrm{Mm}$ and $v_{\mathsmaller{f}} \sim 280\,\mathrm{km/s}$) in the real magnetotail plasma and $\sim 10^{11}$ (corresponding to $L_{\mathsmaller{S}} \sim 1\, \mathrm{Mm}$ and $v_{\mathsmaller{f}} \sim 200\, \mathrm{km/s}$) at the magnetopause. At such a large $Re_{\mathsmaller{m}}$ ($Re_{\mathsmaller{m}} \gg 1$), advection and induction dominate over diffusion, the magnetic field is frozen into the plasma, and the diffusion term in \autoref{eq:eq13} can, in principle, be ignored. 

However, the simulations are not actually devoid of diffusive effects and we are not truly solving ideal MHD because of numerical diffusion. The effective numerical value of $Re_{\mathsmaller{m}}$, are probably closer to $\sim 10^{3}$ which is much less than  the $\sim 10^{11}-10^{15}$ range for the actual magnetospheric plasma estimated above. Numerical diffusion mediates reconnection in the thin current sheets at the center of the magnetotail and also in determining the thickness of the bow shock. Figure 2 illustrates the simulation domain and the Cartesian coordinates used. \textit{(i) Axes:} The X-axis is taken to be along the line connecting the center of the planet and star with the positive direction directed towards the antipode of the substellar point. The Y-axis points toward the north ecliptic pole, which also coincides with the planet's spin-aligned magnetic dipole axis along the N to S zero-tilt magnetic poles. The Z-axis completes the orthogonal right-handed coordinate system. \textit{(ii) Origin:} All  models have the same geometrical configuration, with a non-rotating planet located at the center of the coordinate system and assumed to orbit in the X-Z plane.

The simulations were performed in a Cartesian mesh $(x, y, z)$ having a base grid of $55 \,\times\, 48 \,\times\, 48$ cells and with a base resolution of \commentout{6.6846} $10^{4}\;\text{km}\;(\approx 6.68 \times 10^{-5} \;\text{AU})$ for the magnetized case. We use three additional levels of AMR around the magnetotail and four additional levels around the planet and its atmosphere, allowing a resolution of $10^{4} \big/\, 2^3= 1.25\, \times\, 10^{3}\;\text{km}$ in the tail region and a resolution of $10^{4} \big/\, 2^4 = 6.25\, \times\, 10^{2}\;\text{km}$ in the atmosphere. For the case with a non-magnetic planet, we used a $50 \,\times\, 20 \,\times\, 20$ cells base grid with a $10^{4}\;\text{km}$ resolution and two extra levels (five in total) of AMR that provided a resolution of $312.5\;\text{km}$ in the downtail region of plasma outflow. The planet itself, along with its atmosphere, was resolved with  $\Delta x,\, \Delta y,\, \Delta z \simeq 156.25\;\text{km}$ in this case.

We initialize the entire box outside the  planet's core and atmosphere with a magnetized ambient medium, having  velocity equal to that of the SW. The latter is consistent with an isothermal Parker model \cite{Parker1958} at 1 AU 
associated with a given coronal temperature and dipolar magnetic field for a representative stage of Earth's evolution. Except for the left boundary (Y-Z plane), from which  the SW enters, all other external boundaries of the simulation box have outflow boundary conditions. The simulations presented here evolve until they reach a quasi-steady state. The numerical results are determined to be in a quasi-steady state when the time scale over which the planet loses its atmosphere  $(t_{\mathsmaller{Loss}})$ much exceeds the simulation time $(t_{\mathsmaller{Sim}})$. That is
\begin{equation}
t_{\mathsmaller{Loss}} = \frac{M_{\mathsmaller{Atm}}}{\dot{M}_{\mathsmaller{Atm}}} \gg t_{\mathsmaller{Sim}} 
\label{eq:eq9}
\end{equation}
where $\dot{M}_{\mathsmaller{Atm}}$ represents the mass loss rate of the atmosphere, and $M_{\mathsmaller{Atm}}$ refers to the total atmospheric mass. The total run time is the same for all simulations.

\subsubsection{Solar wind-atmosphere interaction for an unmagnetized Earth}

We also carry out runs for an unmagnetized Earth. Throughout Earth's history, the SW has been both strongly supersonic and super-Alfv$\Acute{\text{e}}$nic. The height above Earth's surface where the wind plasma shocks occur is just outside the standoff location determined from pressure balance (refer to \autoref{eq:eq6}) and is approximately determined by the point where the wind ram pressure equals the atmospheric thermal pressure for an unmagnetized Earth. 

We do not include the Sun in our simulations; rather, we model the direct exposure of the planetary atmosphere to the stellar wind, which leads to its depletion, and focus on the evolution thereafter. The interactions are shown in Figures 5 and ED3, which provide face-on views of the lunar orbital plane. Both the stellar and planetary winds are depicted with arrows and the magnetic fields with solid lines. We note that only the approaching SW is super-sonic, whereas the EW escaping the terrestrial atmosphere and reaching the Moon (an orbit of radius $\sim 60.21 \commentout{60.207}\, R_{\mathsmaller{\oplus}}$, centered on the Earth) is sub-sonic.

When evaluating the average duration of the Moon passing through Earth's tail in one orbit, we take into account the changing Earth-Moon distance over time. The gap between Earth and the Moon has been gradually widening due to tidal dissipation, a process that has been ongoing since the formation of the Earth-Moon system. 

\subsection{Determining the flux of specific ion species}
The model discussed so far has focused on deriving the relative  flux of bulk SW and atmosphere deposition on the moon. The MHD model does not distinguish between specific ion species; to achieve that, we introduce an ionization model to be combined with the calculations from the bulk flows.

Solar extreme ultraviolet (EUV) radiation and soft X-rays shortward of $\lesssim 900$ \r{A} are predominantly absorbed within the Earth's thermosphere, leading to the ionization of various atmospheric species. This not only forms  of the ionosphere but also  photodissociates  molecular components directly from long wavelength photons and via indirect processes driven by the ionization. The most important of these indirect processes is the generation of energetic electrons, which carry  surplus energy imparted by the photoionization. These photoelectrons can further interact with neutral species causing ionization, dissociation, and excitation and  a cascade of elastic and inelastic processes that transfer the initial photon energy into kinetic energy of the various thermospheric or ionospheric constituents. We calculate the orbit-averaged flux for each species  by taking  temporal averages.

\subsubsection{Ionization model}
For our calculation, we assume the upper atmosphere has its base in the ionopause and extends upward to the radius where it merges with the stellar wind, or equivalently, the ambient medium. 

On Earth, at altitudes above $\sim 80$ km, the atmosphere becomes ionized by  solar radiation forming a dense plasma layer  of ionized atmospheric gases and free electrons. Below this altitude, the principal atmospheric constituents  are in essentially  the same relative abundances as at the surface of the Earth, but above they differ. Above the present-day homopause of geodetic altitude  $z_{\mathsmaller{H}} \approx$ 120 km, the atmosphere is in approximate   collisional diffusive equilibrium up to the exobase in which the number density of each constituent species decreases exponentially with altitude. For our assumed isothermal model, this follows the relationship
\begin{equation}
    n_{\mathsmaller{i}}(z) = n_{\mathsmaller{i}}(z_{\mathsmaller{H}})\,e^{-\frac{z - z_{\mathsmaller{H}}}{H_{\mathsmaller{i}}(z)}}
    \label{eq:eq14}
\end{equation}
so that the constituents are distributed vertically based on their individual molecular weights. Here $n_{\mathsmaller{i}}(z_{\mathsmaller{H}})$ is the neutral density of the $i^{th}$ species at the homopause and $H_{\mathsmaller{i}}(z)$ is the neutral gas scale height defined as
\begin{equation}
    H_{\mathsmaller{i}}(z) = \frac{k_{\mathsmaller{B}}\,T_{\mathsmaller{ts}}}{m_{\mathsmaller{i}}\; g(z)}
    \label{eq:eq15}
\end{equation}
where $T_{\mathsmaller{ts}}$ is the thermosphere temperature and $g(z)$ is the altitude-dependent gravity. The radiative transfer calculations of  stellar EUV energy deposition into the upper atmosphere are simplified due to the dominance of  absorption. We also make the following  simplifying assumptions:
\begin{enumerate}[label=\roman*)]
  \item 
  short-wavelength X-rays dominate photoionization,
  \item the atmosphere consists of only those neutral absorbing species that are dominant at the homopause, whose number densities decrease exponentially with altitude, each following its respective scale height, and 
  \item the atmosphere is considered to be plane-parallel and horizontally stratified.
\end{enumerate}

As the photon flux penetrates the atmosphere, it is attenuated by absorption. The solar actinic flux for each wavelength at different atmospheric levels is calculated by applying the Beer-Lambert law layer-by-layer:
\begin{equation}
    \Phi(z,\, \lambda,\, \vartheta_{\mathsmaller{s}}) = \Phi_{\mathsmaller{\infty}}(\lambda)\, e^{-\tau_{\mathsmaller{\lambda}}}
    \label{eq:eq16}
\end{equation}
where the monochromatic optical depth $\tau_{\mathsmaller{\lambda}}$ for a multispecies atmosphere as a function of altitude $z$, wavelength $\lambda$, and the solar zenith angle $\vartheta_{\mathsmaller{s}}$ is given by
\begin{equation}
    \tau_{\mathsmaller{\lambda}} = \tau(z,\, \lambda,\, \vartheta_{\mathsmaller{s}}) = \sec \vartheta_{\mathsmaller{s}} \int\displaylimits_z^\infty \sum\limits_i n_{\mathsmaller{i}}(z)\; \sigma_{\mathsmaller{i}}^{\mathsmaller{(abs)}}(\lambda)\, dz
    \label{eq:eq17}
\end{equation}
where $\sigma_{\mathsmaller{i}}^{\mathsmaller{(abs)}}$ is the absorption cross section. From the estimated neutral density of the constituents and their corresponding absorption cross sections above the homopause, the total ion production rate of each ion species is 
\begin{equation}
    S_{\mathsmaller{tot,\, i}}(z,\, \vartheta_{\mathsmaller{s}}) = n_{\mathsmaller{i}}(z)\,\int\displaylimits_0^{\lambda_{c}} \Phi_{\mathsmaller{\infty}}(\lambda)\, e^{-\tau_{\mathsmaller{\lambda}}} \; \sigma_{\mathsmaller{i}}^{\mathsmaller{(ion)}}(\lambda)\, d\lambda
    \label{eq:eq18}
\end{equation}

We consider the exobase to be the lower boundary at which various mass-loss processes kick in. It is defined as the altitude where the mean free path of the atmospheric particles becomes large compared to the particle-specific atmospheric scale height, and collisions become negligible (Knudsen number $\gtrsim 1$), so that atmospheric constituents with velocities exceeding the  escape velocity can efficiently escape from the planet. Above the exobase, the electric field induced by the magnetized SW flow and the intrinsic dipole field during the planet's magnetized phase accelerates the terrestrial ions, which move away from Earth along with the redirected stellar wind flow around the planet. This pick-up process and the escape mechanism can efficiently remove terrestrial ions produced above the exobase.

To calculate the escape flux of different species that result from ion pick-up above the exobase, we first determine the ion production rates above the homopause. We employ the NRLMSISE-00 model \cite{Picone2002}  as an input. This is an empirical model characterizing the neutral densities of the constituents in the terrestrial atmosphere. The EUVAC flux model \cite{Richards1994}, driven by the variations of the 10.7 cm solar radio flux, is used for the reference solar spectrum. This model is based on the measured F74113 solar EUV reference spectrum \cite{Hinteregger1981} and provides fluxes in 37 wavelength intervals, covering a range of 50 to 1050 \r{A}. In the EUVAC model, the solar flux is obtained from the reference spectra using the following scaling relation:
\begin{equation}
    \Phi_{\mathsmaller{\infty}}(\lambda) = \frac{\Phi_{\mathsmaller{ref}}(\lambda)}{d^2}\, \left[1 + A_{\mathsmaller{\lambda}} \left(P - 80\right)\right]
    \label{eq:eq19}
\end{equation}
where $\Phi_{\mathsmaller{ref}}(\lambda)$ is the binned reference flux, $d$ is the distance of the planet from its host star in AU, $A_{\mathsmaller{\lambda}}$ is the scaling factor for each wavelength interval,  and $ P = \left(F_{\mathsmaller{10.7}} + \left\langle F_{\mathsmaller{10.7}} \right\rangle\right) / 2$ accounts for solar activity variations. Here $F_{\mathsmaller{10.7}}$ is the solar radio flux index and $\left\langle F_{\mathsmaller{10.7}} \right\rangle$ is its 81-day centered mean. 
We take $F_{\mathsmaller{10.7}}=\left\langle F_{\mathsmaller{10.7}} \right\rangle=150\, \mathrm{W\, m^{-2}\, Hz^{-1}},$ representing a typical EUV flux and solar activity. The ion production rates are then computed from the photo- and electron-impact ionization of neutrals by the incident stellar radiation (Figure ED2).

Assuming that ions produced above the exobase are picked up by the SW and escape from the planet, both in the magnetized and unmagnetized cases, we estimated the total ion escape rate by integrating the ion production rate over the dayside planet surface area above the exobase.

The picked-up terrestrial atmospheric ions flow with the SW and are carried anti-sunward down the magnetotail within the cross section of the magnetosphere. The roughly paraboloid-shaped magnetopause has an area of about \commentout{1.061501} $1.06 \times 10^{11} \, \mathrm{km^2}$ for magnetized Earth and \commentout{6.9877523} $6.99  \times 10^{9} \, \mathrm{km^2}$ in the unmagnetized case at the lunar orbit apogee location. Relative escape fluxes of different ion species in the nightside tail region are calculated by normalizing the ion escape rates by the magnetotail cross-sectional area. The estimated escape flux of each constituent as a function of the geodetic altitude is displayed in Figure 4.

\subsubsection{Unmagnetized early Archean atmosphere composition}

Standard solar evolution models suggest that the Sun’s total luminosity has increased by about 30–35\% over the past $\sim 4.6$ billion years of geologic time \cite{Gilliland1989,Gough1981}, meaning that the Earth's surface, with a present-day atmosphere, similar orbit, and comparable level of greenhouse forcing, would have below-freezing temperatures \cite{Sagan1972,Goldblatt2011,Feulner2012}. In contrast, geological evidence indicates that the early terrestrial climate was warm and that liquid water was ubiquitously present at the surface from as early as the Hadean epoch through the Archean era \cite{Catling2020,Foriel2004}. The most widely accepted hypothesis that reconciles this apparent faint young Sun paradox involves an enhanced abundance of greenhouse gas to compensate for the lower intensity of primordial solar radiation \cite{Tarduno2014}. Such forcing could occur if the homosphere of Archean Earth had more $\mathrm{CO}_{2}$ or produced such gases (e.g., $\mathrm{N_{2}O}$) from increased thermospheric heating caused by the more intense XUV radiation from the young Sun \cite{Herwartz2021,Goldblatt2009,Johnstone2021}.

We utilize the model of the early Archean homosphere by Johnstone et al. \cite{Johnstone2021}, which is composed of equal parts $\mathrm{CO}_{2}$ and $\mathrm{N}_{2}$ (49.95\% each). The average Archean solar XUV flux is  an order of magnitude higher than the present-day value \cite{Judge2003,Johnstone2021}. We therefore correspondingly increase the solar activity proxy ($P$ in \autoref{eq:eq19}) to scale the measured F74113 solar EUV reference and obtain the EUVAC solar irradiance spectra for a young Sun. The hydrogen abundance of the Archaean lower atmosphere, including its influence on Earth's early climate as an indirect greenhouse gas, is uncertain \cite{Wordsworth2013,Kasting2014}. But the Archean atmosphere is unlikely to have been $\mathrm{H}_{2}$-rich, considering its low outgassing from Earth's crust and its microbial consumption, resulting in a low partial pressure \cite{Kadoya2019,Catling2020}. Following Kasting et al. \cite{Kasting1993}, we adopt low $\mathrm{H}_{2}$ mixing ratios of the order of 0.1\% by volume. Study of the composition of nitrogen and argon isotopes in fluid inclusions encapsulated in Archean hydrothermal quartz indicates $\mathrm{N}_{2}$\big/Ar ratios of air-saturated water (ASW) comparable to those of present-day  \cite{Marty2013}. We therefore assumed a constant concentration of atmospheric Ar at the Archean homopause that matches the current ASW value of $1.0-1.3 \times 10^{4}$. Ne also exhibits a similar trend; the atmospheric Ne composition during the early Earth is attributed to a mixing of mantle gases and chondritic gases delivered in a late veneer phase, and did not undergo much variation thereafter \cite{Sandrine2018}. We choose the atmospheric Ne/Ar ratio to be 0.051, akin to the present-day atmospheric value \cite{Avice2018,Sandrine2018}. Assuming complete degassing from the continental crust, an analogously steady He concentration, originating from the decay of radioactive elements and seeping  into the atmosphere \cite{Arakawa2012,Vardiman1986}, also applies. These values are also consistent with the estimates of noble gas isotopic abundances in the early Earth's atmosphere, which entered the atmosphere directly through meteorite or asteroid impact, or degassed from volatiles delivered to Earth after the loss of a primary atmosphere \cite{Zahnle2010}.

\subsubsection{Combining ionization model with total flux to predict lunar isotope ratios}

To  determine the relative elemental abundance of different major species in the SW and Earth's atmosphere that are transported to the Moon, we combine results from our 3D-MHD model with the ionization  model. We compute the impinging relative fluxes of H, N, and light noble gases by normalizing the flux of each species by the total escape flux of all constituents obtained from  photoionization. We then scale these values so that the total escape flux equals the orbit-averaged planetary flux computed from the 3D-MHD simulations.

Such a scaling of the escaping flux is possible even if the plasma is  collisionless, which it actually is above the exobase; then, the viscosity is anisotropic, with a lower value across the field lines due to the mean free path being approximately the same as that of the gyroradius of ions.  Thus the plasma is trapped by field lines perpendicular to the flow but moves more freely parallel to the magnetic field. On the nightside, the terrestrial ions are trapped by the magnetic field of the SW into the wind flow and move downtail with the bulk velocity because the gyro-radius is small compared to the macroscopic system scales.  In particular, the characteristic gyro radius of for a typical magnetotail ion is given by 
\begin{equation}
    r_{\mathsmaller{g},\,\mathsmaller{i}} = 10.44 \,\mathrm{km} \left( \frac{\mu_{\mathsmaller{i}}}{Z_{\mathsmaller{i}}}\right) \left(\frac{v_{\mathsmaller{Bulk}}}{1\, \mathrm{km / s}}\right) \left(\frac{B_{\mathsmaller{SW}}}{1\, \mathrm{nT}}\right)^{-1}
    \label{eq:eq30}
\end{equation}
where $Z_{\mathsmaller{i}}$ is the ion charge state and $\mu_{\mathsmaller{i}}$ is the normalization factor used to normalize ion mass by the proton mass. For a proton advecting with the bulk flow velocity $(\sim 268\,\,\mathrm{km/s}\;\, \text{in Case -- \text{I}})$ in the magnetotail, the gyroradius is $\sim 560$ km, indicating that it is roughly of the same order as the scale height used in our MHD simulation $(\sim 10^3\,\, \mathrm{km})$, thereby justifying  our  approximation of using a combined MHD-ionization model for all larger scales.

\subsubsection{Binary mixing model to infer the non-solar component flux}

We apply a binary mixing model \cite{Langmuir1978} to estimate contributions from the SW and NSC, using their distinct isotope ratios and absolute abundances of different elements. We generate the theoretical binary mixing models by varying the proportions of each end member in incremental steps. The mixing equation for the two end-members, NSC and SW, is the average of the isotopic ratios for an element in the end-members, weighted by their respective mass fractions in the mixture. If NSC and SW components contain different concentrations of element $\mathcal{X}$, and if $x/X$ represents the isotope ratio of $\mathcal{X}$, then the binary mixing equation \cite{Faure1986} can be expressed as
\begin{equation}
    \left( \dfrac{x}{X}\right)_{\mathsmaller{M}} = \dfrac{\left( \dfrac{x}{X}\right)_{\mathsmaller{SW}} \,\mathcal{X}_{\mathsmaller{SW}} \, \,\text{\textcursive{\slshape \footnotesize f\,}$_{\mathsmaller{SW}}$} + \left( \dfrac{x}{X}\right)_{\mathsmaller{NSC}} \,\mathcal{X}_{\mathsmaller{NSC}} \, \,\text{\textcursive{\slshape \footnotesize f\;}$_{\mathsmaller{NSC}}$}}{\mathcal{X}_{\mathsmaller{SW}} \, \,\text{\textcursive{\slshape \footnotesize f\,}$_{\mathsmaller{SW}}$} + \mathcal{X}_{\mathsmaller{NSC}} \, \,\text{\textcursive{\slshape \footnotesize f\;}$_{\mathsmaller{NSC}}$}}
    \label{eq:eq31}
\end{equation}
where \textcursive{\slshape \footnotesize f\;}$_{\mathsmaller{SW}}$ and \textcursive{\slshape \footnotesize f\;}$_{\mathsmaller{NSC}}$  stand for the mass fractions of SW and NSC subject to the constraint \textcursive{\slshape \footnotesize f\;}$_{\mathsmaller{SW}}$ + \textcursive{\slshape \footnotesize f\;}$_{\mathsmaller{NSC}}$ $ = 1$, and
$\left( \frac{x}{X}\right)_{\mathsmaller{M}}$ denotes the isotope ratio of $X$ in a mixture  of SW and NSC, while $\mathcal{X}_{\mathsmaller{SW}}$ and $\mathcal{X}_{\mathsmaller{NSC}}$ are the respective concentrations of $\mathcal{X}$ in the SW and NSC. Thus, $\left( \frac{x}{X}\right)_{\mathsmaller{SW}} \mathcal{X}_{\mathsmaller{SW}}$ is the species-specific isotope fraction of the total elemental abundance. Similarly, the isotope composition of element $\mathcal{Y}$ in the suite of binary mixture of SW and NSC is given by:
\begin{equation}
    \left( \dfrac{y}{Y}\right)_{\mathsmaller{M}} = \dfrac{\left( \dfrac{y}{Y}\right)_{\mathsmaller{SW}} \,\mathcal{Y}_{\mathsmaller{SW}} \, \,\text{\textcursive{\slshape \footnotesize f\,}$_{\mathsmaller{SW}}$} + \left( \dfrac{y}{Y}\right)_{\mathsmaller{NSC}} \,\mathcal{Y}_{\mathsmaller{NSC}} \, \,\text{\textcursive{\slshape \footnotesize f\;}$_{\mathsmaller{NSC}}$}}{\mathcal{Y}_{\mathsmaller{SW}} \, \,\text{\textcursive{\slshape \footnotesize f\,}$_{\mathsmaller{SW}}$} + \mathcal{Y}_{\mathsmaller{NSC}} \, \,\text{\textcursive{\slshape \footnotesize f\;}$_{\mathsmaller{NSC}}$}}
    \label{eq:eq32}
\end{equation}
For example, in terms of the ordinate of the $\mathrm{{}^{15}N / {}^{14}N} - \mathrm{D / H}$ mixing diagram (Figure 6), the equation above can be represented as:
\begin{equation}
    \left( \dfrac{\mathrm{{}^{15}N}}{\mathrm{{}^{14}N}}\right)_{\mathsmaller{M}} = \dfrac{\left( \dfrac{\mathrm{{}^{15}N}}{\mathrm{{}^{14}N}}\right)_{\mathsmaller{SW}} \,\mathrm{N}_{\mathsmaller{SW}} \, \,\text{\textcursive{\slshape \footnotesize f\,}$_{\mathsmaller{SW}}$} + \left( \dfrac{\mathrm{{}^{15}N}}{\mathrm{{}^{14}N}}\right)_{\mathsmaller{NSC}} \,\mathrm{N}_{\mathsmaller{NSC}} \, \,\text{\textcursive{\slshape \footnotesize f\;}$_{\mathsmaller{NSC}}$}}{\mathrm{N}_{\mathsmaller{SW}} \, \,\text{\textcursive{\slshape \footnotesize f\,}$_{\mathsmaller{SW}}$} + \mathrm{N}_{\mathsmaller{NSC}} \, \,\text{\textcursive{\slshape \footnotesize f\;}$_{\mathsmaller{NSC}}$}}
    \label{eq:eq33}
\end{equation}
 The mass fraction of the NSC component (\textcursive{\slshape \footnotesize f\;}$_{\mathsmaller{NSC}}$) in the binary mixture of solar and Earth winds can be calculated if the isotopic compositions and elemental abundances for both end-members as well as the isotopic composition of the presumed mixture (represented by the above \hyperref[eq:eq31]{mixing equation}) are known:
\begin{equation}
    \text{\textcursive{\slshape \footnotesize f\;}$_{\mathsmaller{NSC}}$} = \frac{\mathcal{X}_{\mathsmaller{SW}} \, \left[\left( \dfrac{x}{X}\right)_{\mathsmaller{SW}} - \left( \dfrac{x}{X}\right)_{\mathsmaller{M}} \right]}{\left( \dfrac{x}{X}\right)_{\mathsmaller{M}} \left(\mathcal{X}_{\mathsmaller{NSC}} - \mathcal{X}_{\mathsmaller{SW}}\right) \, - \, \left( \dfrac{x}{X}\right)_{\mathsmaller{NSC}}\,\mathcal{X}_{\mathsmaller{NSC}}\,+\, \left( \dfrac{x}{X}\right)_{\mathsmaller{SW}}\,\mathcal{X}_{\mathsmaller{SW}}}
    \label{eq:eq29}
\end{equation}
The exact analytical solution for a binary mixture of two elements is obtained by combining \hyperref[eq:eq31]{Equations 29} and \ref{eq:eq32} and eliminating \textcursive{\slshape \footnotesize f\,}$_{\mathsmaller{SW}}$. The non-solar mass fraction of an element (say $\mathcal{X}$) is the weight fraction of that element in the two end-member mixture. Mixing between the end members generally forms hyperbolic relationships \cite{Vollmer1976,Langmuir1978} of the form
\begin{equation}
    a\,\left( \dfrac{x}{X}\right)_{\mathsmaller{M}}\,+\,b\,\left( \dfrac{x}{X}\right)_{\mathsmaller{M}} \left( \dfrac{y}{Y}\right)_{\mathsmaller{M}} \, + \,c\,\left( \dfrac{y}{Y}\right)_{\mathsmaller{M}}\,+\,d = 0
    \label{eq:eq27}
\end{equation} 
Thus, the mixing equation for the $\mathrm{{}^{15}N / {}^{14}N} - \mathrm{D / H}$ system (Figure 6) can be expressed as 
\begin{equation}
    a\,\left( \dfrac{\mathrm{{}^{15}N}}{\mathrm{{}^{14}N}}\right)_{\mathsmaller{M}}\,+\,b\,\left( \dfrac{\mathrm{{}^{15}N}}{\mathrm{{}^{14}N}}\right)_{\mathsmaller{M}} \left( \dfrac{\mathrm{D}}{\mathrm{H}}\right)_{\mathsmaller{M}} \, + \,c\,\left( \dfrac{\mathrm{D}}{\mathrm{H}}\right)_{\mathsmaller{M}}\,+\,d = 0
    \label{eq:eq34}
\end{equation}
The coefficients of the binary mixing equation (\autoref{eq:eq27}) are:
\begin{equation}
   \begin{aligned}
    a &= \left( \dfrac{y}{Y}\right)_{\mathsmaller{NSC}}\,\mathcal{Y}_{\mathsmaller{NSC}}\,\mathcal{X}_{\mathsmaller{SW}} - \left( \dfrac{y}{Y}\right)_{\mathsmaller{SW}}\,\mathcal{Y}_{\mathsmaller{SW}}\,\mathcal{X}_{\mathsmaller{NSC}}  \\[5pt]
    b &= \mathcal{Y}_{\mathsmaller{SW}}\,\mathcal{X}_{\mathsmaller{NSC}} - \mathcal{Y}_{\mathsmaller{NSC}}\,\mathcal{X}_{\mathsmaller{SW}}  \\[5pt]
    c &= \left( \dfrac{x}{X}\right)_{\mathsmaller{SW}}\,\mathcal{Y}_{\mathsmaller{NSC}}\,\mathcal{X}_{\mathsmaller{SW}} - \left( \dfrac{x}{X}\right)_{\mathsmaller{NSC}}\,\mathcal{Y}_{\mathsmaller{SW}}\,\mathcal{X}_{\mathsmaller{NSC}}  \\[5pt]
    d &= \left( \dfrac{y}{Y}\right)_{\mathsmaller{SW}}\, \left( \dfrac{x}{X}\right)_{\mathsmaller{NSC}}\,\mathcal{Y}_{\mathsmaller{SW}}\,\mathcal{X}_{\mathsmaller{NSC}} - \left( \dfrac{y}{Y}\right)_{\mathsmaller{NSC}}\, \left( \dfrac{x}{X}\right)_{\mathsmaller{SW}}\,\mathcal{Y}_{\mathsmaller{NSC}}\,\mathcal{X}_{\mathsmaller{SW}} \\[5pt]
  \end{aligned}
\label{eq:eq28}
\end{equation}
where, as briefly described earlier, $\mathcal{X}_{\mathsmaller{SW}}$, $\mathcal{X}_{\mathsmaller{NSC}}$, $\mathcal{Y}_{\mathsmaller{SW}}$, and $\mathcal{Y}_{\mathsmaller{NSC}}$ are elemental abundances of   the end members, i.e. pure SW or pure NSC, and $\left( \frac{x}{X}\right)_{\mathsmaller{SW}}$, $\left( \frac{y}{Y}\right)_{\mathsmaller{SW}}$,
 $\left( \frac{x}{X}\right)_{\mathsmaller{NSC}}$ and $\left( \frac{y}{Y}\right)_{\mathsmaller{NSC}}$ are the end member isotope ratios.

A curvature function $K_{\mathsmaller{c}}$ characterizes the shapes of the theoretical mixing hyperbolas and is used to diagnose the degree of mixing between the end members and estimate the fraction of NSC in the lunar samples. This $K_{\mathsmaller{c}}$ depends on the $b$ coefficient and is defined as $\mathrm{[ N / H ]_{\mathsmaller{\mathrm{SW}}}\,\big/ \, [ N / H ]_{\mathsmaller{\mathrm{EW}}}}$ for the N-H isotope plot, $\mathrm{[ He / Ne ]_{\mathsmaller{\mathrm{SW}}}\,\big/ \, [ He / Ne ]_{\mathsmaller{\mathrm{EW}}}}$ for the He-Ne isotope plot, and $\mathrm{[ He / Ar ]_{\mathsmaller{\mathrm{SW}}}\,\big/ \, [ He / Ar ]_{\mathsmaller{\mathrm{EW}}}}$ for the He-Ar isotope plot, respectively.

Along each mixing curve, which corresponds to a particular exobase height, from one end-member (say NSC) to the other (SW), the mass fraction of the starting end-member decreases (here, \textcursive{\slshape \footnotesize f\;}$_{\mathsmaller{NSC}}$), and the mass fraction of the terminating end-member increases (\textcursive{\slshape \footnotesize f\,}$_{\mathsmaller{SW}}$), maintaining the sum of fractions to be unity along the path.  Mixing between the NSC and SW with different isotopic compositions results in isotope fraction values, $\left( \frac{y}{Y}\right)_{\mathsmaller{M}} \text{and} \left( \frac{x}{X}\right)_{\mathsmaller{M}}$, that are intermediate between the two end-members. For example, if our mixing diagram represents $\mathrm{{}^{15}N / {}^{14}N}$ versus $\mathrm{D / H}$ (Figure 6), then the mass fraction of NSC (\textcursive{\slshape \footnotesize f\;}$_{\mathsmaller{NSC}}$), which comprises isotopes of both the elements nitrogen and hydrogen, is unity at the terrestrial component point (denoted by a navy square) in the plot. The value of \textcursive{\slshape \footnotesize f\;}$_{\mathsmaller{NSC}}$ decreases to zero along the mixing trajectory non-uniformly as one progresses towards the SW component point (denoted by an orange square). 
Conversely, \textcursive{\slshape \footnotesize f\,}$_{\mathsmaller{SW}}$ increases along the path. 

The non-uniform  decrement of \textcursive{\slshape \footnotesize f\;}$_{\mathsmaller{NSC}}$ or increment of \textcursive{\slshape \footnotesize f\,}$_{\mathsmaller{SW}}$ per unit length along a curve is due to the curvature of the rectangular hyperbolas. The extent and direction of the curvature are functions of the concentration contrast between the quantities in the denominator of each isotope ratio under consideration ($K_{\mathsmaller{c}}$), representing the relative isotopic abundances of the end-members in the mixture. $K_{\mathsmaller{c}}$ varies between zero and infinity. Only in the special case where the ratio of the two denominators in the two end-members is equal ($K_{\mathsmaller{c}} = 1$) is the mixing curve a straight line. As $K_{\mathsmaller{c}}$ becomes progressively greater than or less than unity, the hyperbolic concavity becomes more pronounced. In the $\mathrm{{}^{15}N / {}^{14}N}$ - $\mathrm{D / H}$ mixing plot (Figure 6), the hyperbolic curves for exobase heights of $221$ km (solid line), $275$ km (dash-dot line), $300$ km (dashed line), and $325$ km (densely dotted line) display a concave-down shape, signifying $K_{\mathsmaller{c}} \in (0,\,1)$, whereas the profiles corresponding to exobase altitudes of $350$ km (dotted line) and $400$ km (dash-dot-dotted line) exhibit concave-up behavior, suggesting $K_{\mathsmaller{c}} > 1$. When all data points lie along a single mixing curve, this indicates consistency with a specific exobase choice with variation in \textcursive{\slshape \footnotesize f\;}$_{\mathsmaller{NSC}}$ (or \textcursive{\slshape \footnotesize f\,}$_{\mathsmaller{SW}}$).

\clearpage


\begin{figure}
    \centering
	\includegraphics[scale=0.5]{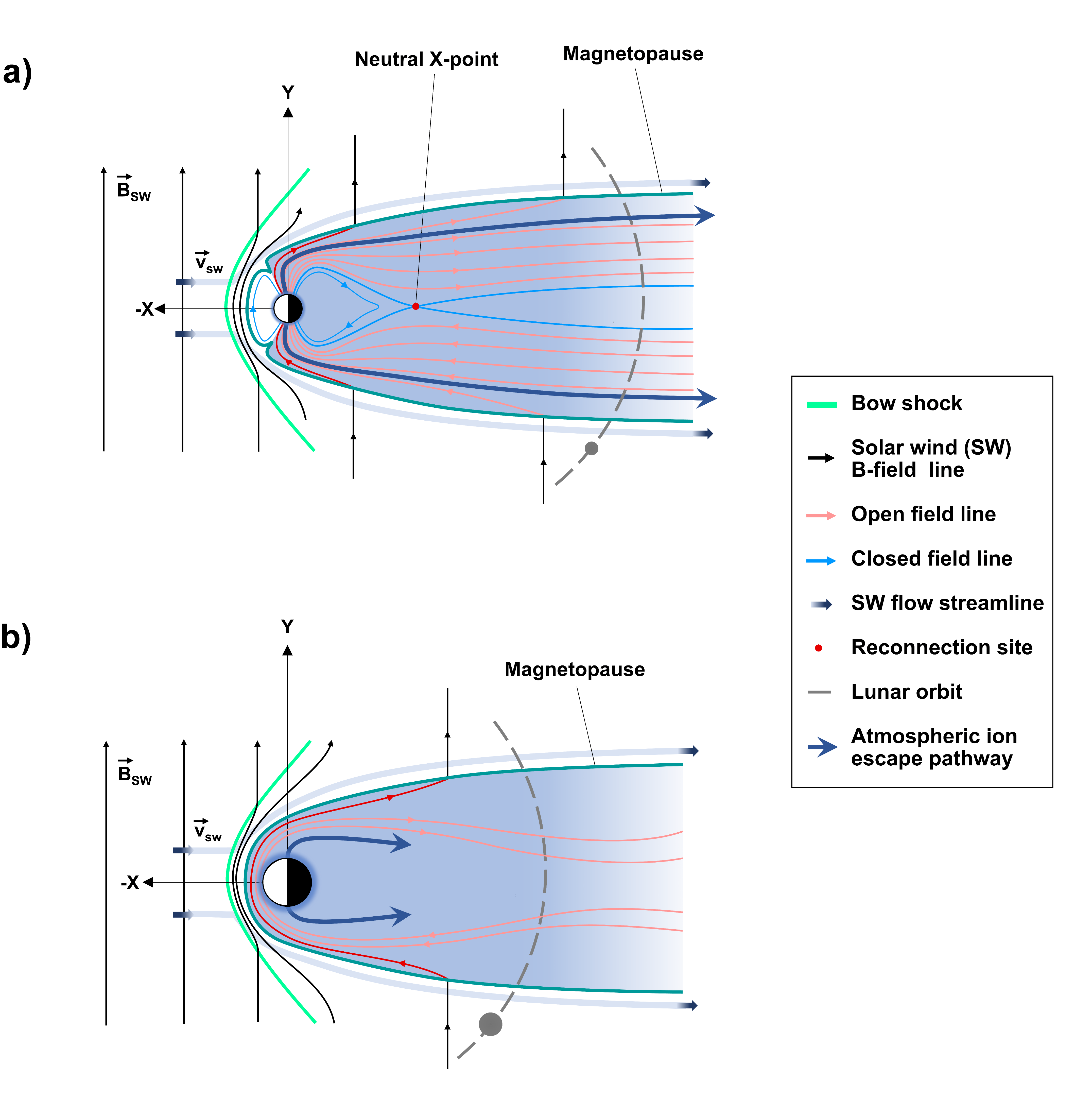}
    \caption{\textbf{Effect of geodynamo on terrestrial atmospheric ion escape.} \textit{Top [panel (a)]}: Schematic representation (not to scale) of the solar wind magnetic field and intrinsic dipolar field in the magnetosphere and escaping plasma outflow from present-day upper terrestrial ionosphere [modified after Hultqvist et al. \cite{Hultqvist1999}]. The bluish-gray color depicts the region where atmospheric ions are commonly observed, and the Prussian blue lines highlight their escape pathways. The magnetic fields are represented by the black lines, while the Moon's orbit is indicated by the dashed gray line. \textit{Bottom [panel (b)]}: Illustration (not to scale) of atmospheric ion escape in an induced magnetosphere during a hypothetical phase when Earth lacked a core dynamo. The atmosphere acts as a conductive barrier to the supersonic and super-Alfv$\Acute{\text{e}}$nic SW, deflecting it around due to the magnetic pressure generated by induced currents, resulting in the formation of a plasmapause boundary on the dayside and a magnetotail on the nightside. The black lines represent SW magnetic fields, while the geraldine lines show the field induced by the interaction of the wind with the planetary atmosphere.
    }
    \label{fig:1}
\end{figure}

\clearpage


\begin{figure}
    \centering
	\includegraphics[scale=0.40]{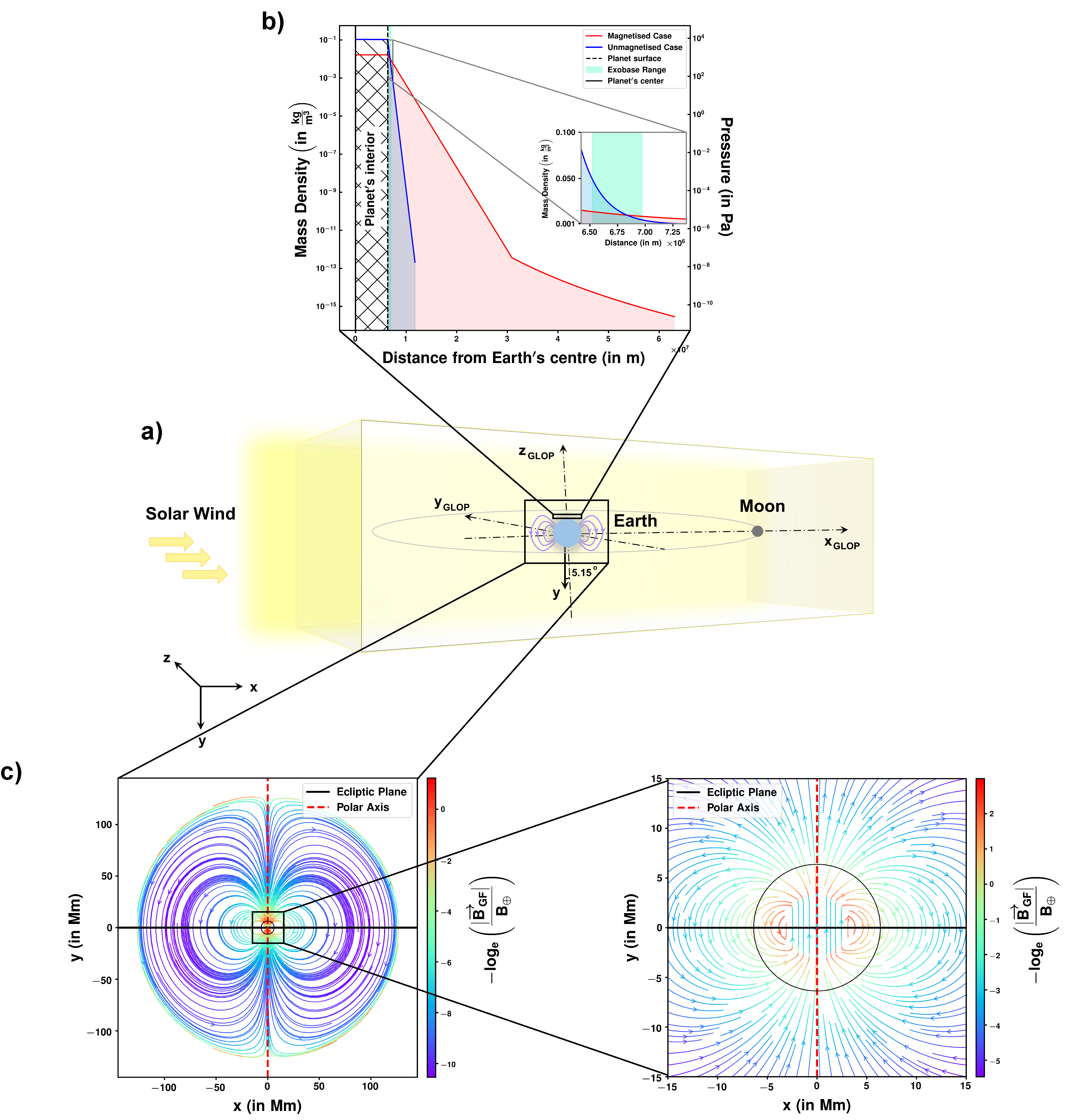}
    \caption{Description continued on the following page.}
    \label{fig:2}
\end{figure}

\begin{figure}
    \ContinuedFloat
    \caption{\textbf{Schematic representation of the simulation framework for solar wind-Earth atmosphere interactions, incorporating terrestrial atmosphere profiles and the geodynamo model used in this study, along with the coordinate frames employed for analysis.} \textit{(a)} Illustrated is the MHD computational domain used for modeling solar wind interaction with the terrestrial atmosphere. Both the Earth and the Moon lie within the simulation box while the solar wind impinges on the left boundary (Y-Z plane) with a velocity normal to the plane. The incoming wind is magnetized, and the field lines are aligned to the Y-axis. We used the Geocentric-Lunar Orbital Plane coordinate frame, which is tilted by the same amount as the Moon's orbital plane, to denote the changing phase angle of the Moon in its orbit and to compute the solar and Earth wind mass fluxes over a complete lunation. \textit{(b)} Static isothermal atmosphere at the start of a run (Case -- \text{I}) for a present-day magnetized (refer to \autoref{eq:eqAtmMag}) and an Archean unmagnetized Earth (see \autoref{eq:eqAtmUnmag}). The interior of the planet is also assumed to be isothermal. The red and blue shaded regions represent the fixed total mass in the two subcases, while the turquoise region illustrates the range of exobase heights used in the binary mixing models. The kink in the red line denotes the $\upbeta_{\mathsmaller{p}} = 1$ surface, outside of which the magnetic field governs the plasma outflow. \textit{(c)} \textit{Left Panel}: Cross section of the planetary magnetic field $(\protect\vec{B}_{\mathsmaller{GF}})$ in the noon-midnight meridian plane used in the simulations at the beginning of the runs ($t = 0$). The color contour shows the log-normalized strength of the B-field $(\mathrm{B}_{\mathsmaller{\oplus}} = 3.12 \times 10^{4} \text{ nT})$, and the arrows denote the direction of field lines. We use a cutoff boundary of $20\,R_{\mathsmaller{\oplus}}$ beyond which the influence of the intrinsic field is essentially insignificant. The horizontal axis aligns with the ecliptic plane (marked by the solid black line), while the polar axis (indicated by the dashed red line) points from dusk to dawn. \textit{Right Panel}: The zoomed-in slice of the field structure in the interior of the planet. The solid, thick circle represents the planet's surface.}
\end{figure}

\clearpage



\begin{figure}
    \centering
	\includegraphics[scale=0.43]{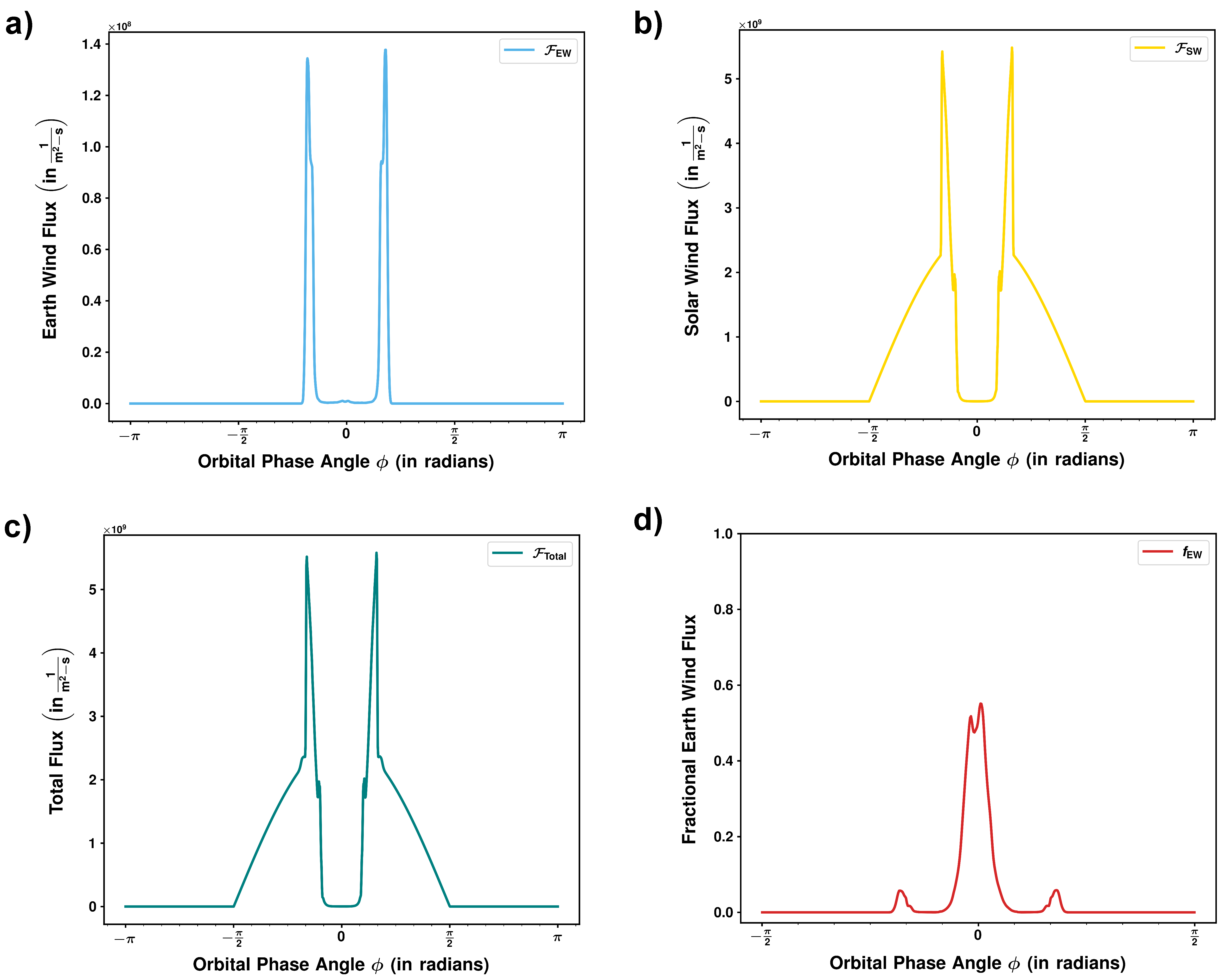}
    \caption{\textbf{Orbital variations of species-total mass fluxes impinging on the lunar surface in the magnetized Earth case.} Angular distribution of the \textit{(a)} planet atmosphere flux, \textit{(b)} solar wind flux, \textit{(c)} total flux (aggregate flux of the terrestrial atmosphere and solar wind), and \textit{(d)} fractional contribution of planetary wind to the total flux over a full lunation. At phase angles $-\pi,\, \pi$ rad, the Moon is in the dayside upstream solar wind at the sub-solar line of Earth. The double-horned structures in the flux profiles result from shocked material located between the magnetosheath and the magnetotail lobe.}
    \label{fig:3}
\end{figure}

\clearpage



\begin{figure}
    \centering
	\includegraphics[scale=0.5]{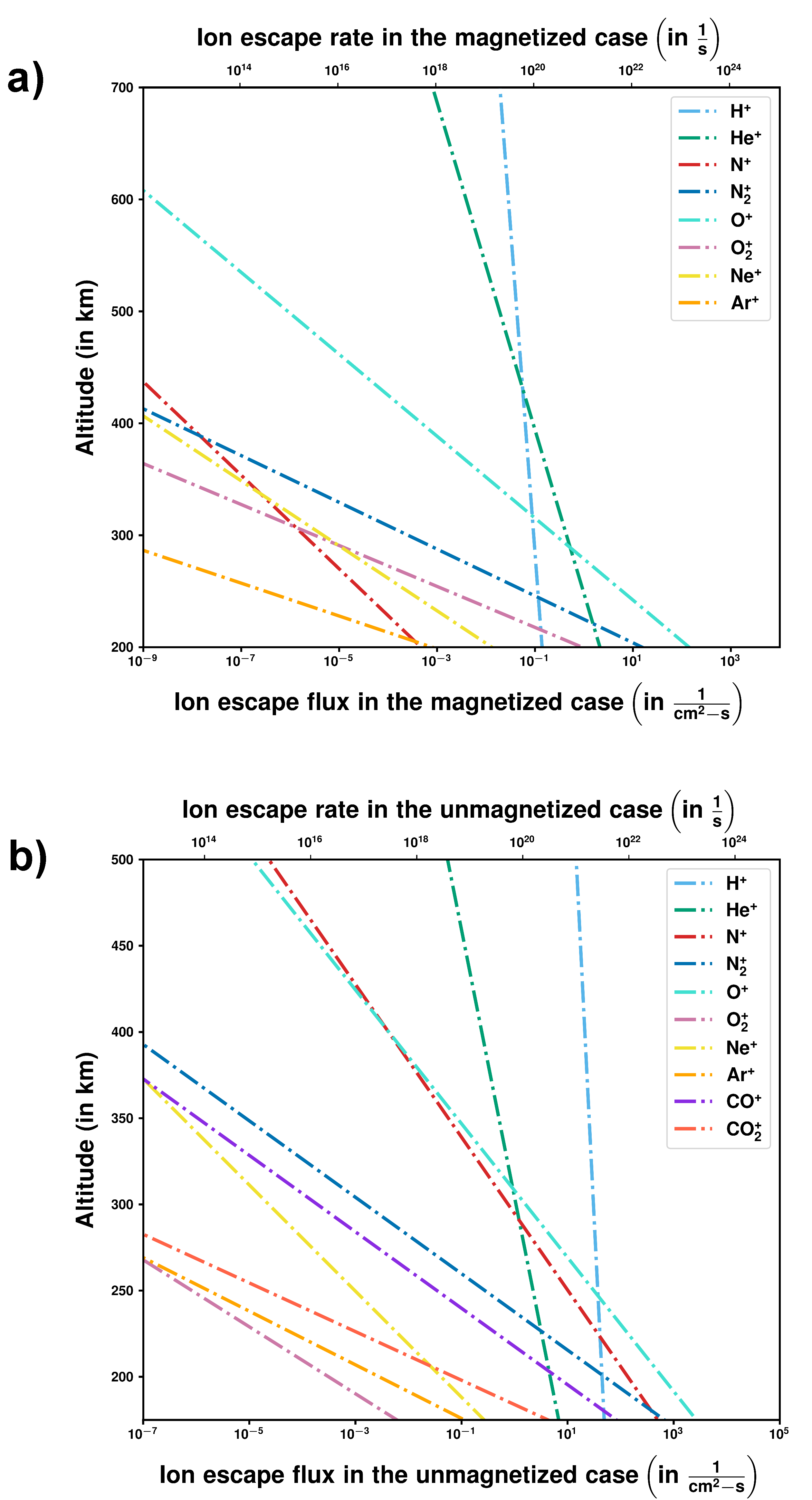}
    \caption{\textbf{Altitude profiles of the escape rates of volatile species due to the collisionless nature of the plasma for present-day Earth (with a geodynamo) and Archean Earth (lacking a geodynamo).} \textit{Top panel (a)}: Ion escape rates (secondary horizontal axis, top $x$-axis) of different constituent species for magnetized Earth calculated using the NRLMSISE00 model of the present terrestrial neutral upper atmosphere and ion production rates for normal solar activity condition (as described in Figure ED2) assuming that ions produced above the exobase escape from the planet \cite{Ozima2005}. Escape fluxes (primary horizontal axis, bottom $x$-axis) of ions in the nightside tail region is computed by normalizing the escape rates by the magnetotail cross-sectional area at the apogee point of the Moon's orbit. \textit{Bottom panel (b)}: Same as \textit{panel (a)}, but for unmagnetized Earth with Archean terrestrial upper atmosphere, representing primordial Earth case.}
    \label{fig:4}
\end{figure}

\clearpage


\newgeometry{top=2cm, bottom=1.7cm, left=0.85cm, right=0.75cm}

\begin{figure*}
    \centering
	\includegraphics[scale=0.43]{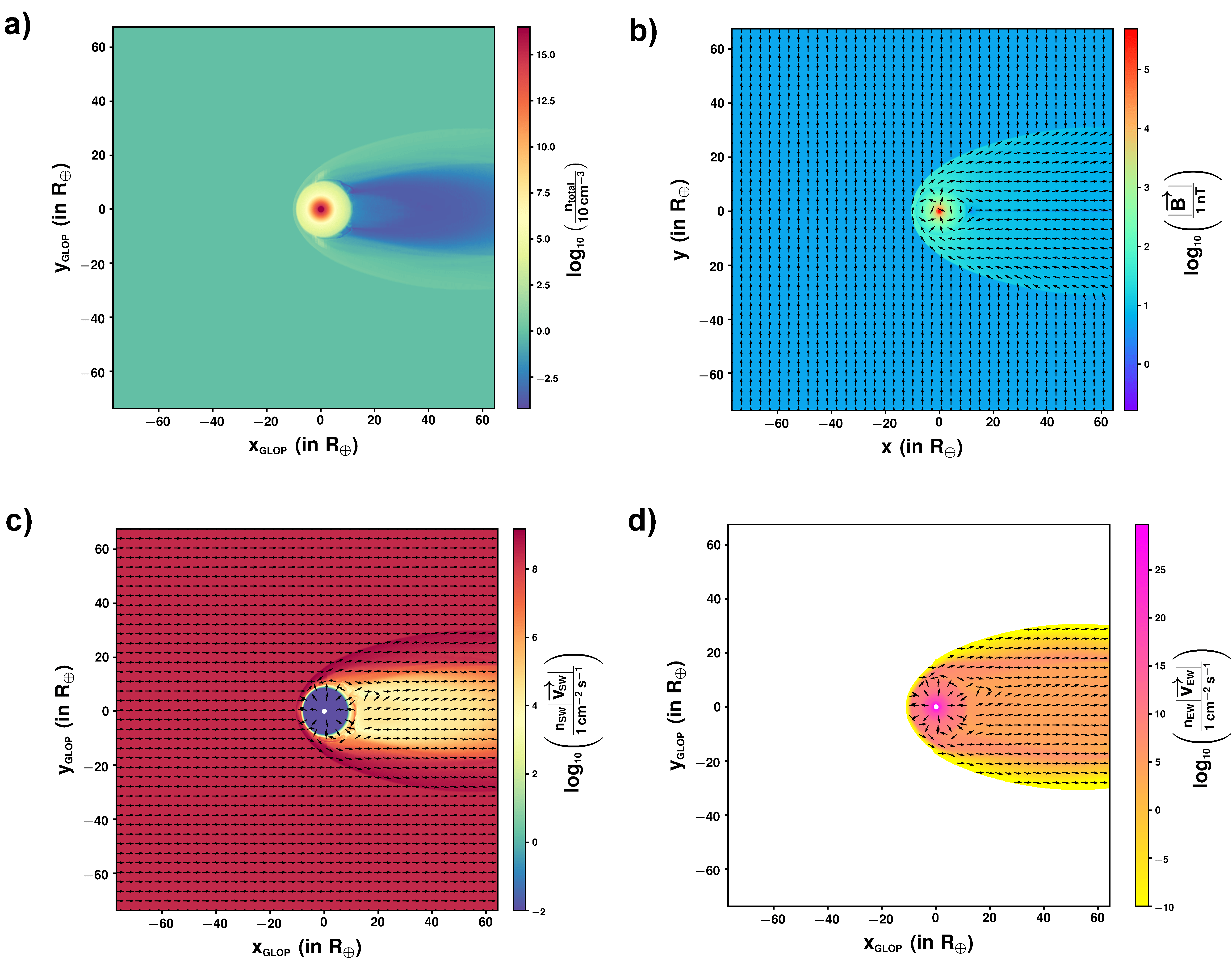}
    \caption{\textbf{Results from our 3D-MHD simulation model for the case in which the planet possesses an intrinsic dynamo (Case -- \text{I}).} \textit{Panel (a)}: Total number density plot depicting the large-scale structure of the wind-atmosphere interaction upto the lunar orbit in the quasi-steady state. \textit{Panel (b)}: Polar-cut view of the topology of the planet's magnetosphere when it is impinged by magnetized solar wind. Quivers represent the magnetic field. \textit{Panel (c)}: 2D slice of the normalized solar wind mass flux in the lunar orbital plane. The flux drops by $\sim 4$ orders of magnitude inside the magnetotail compared to the shocked solar wind region between the bow shock and the magnetopause boundary, where it exhibits a sharp gradient. \textit{Panel (d)}: Terrestrial atmosphere mass flux in the magnetotail region. The quiver arrows denote the plane-projected bulk velocity component indicating the escaping pathways. The filled white circle representing the planet's core is not included in the passive advective tracers.
    }
    \label{fig:5}
\end{figure*}
\restoregeometry
\clearpage



\begin{figure}
    \centering
	\includegraphics[scale=0.5]{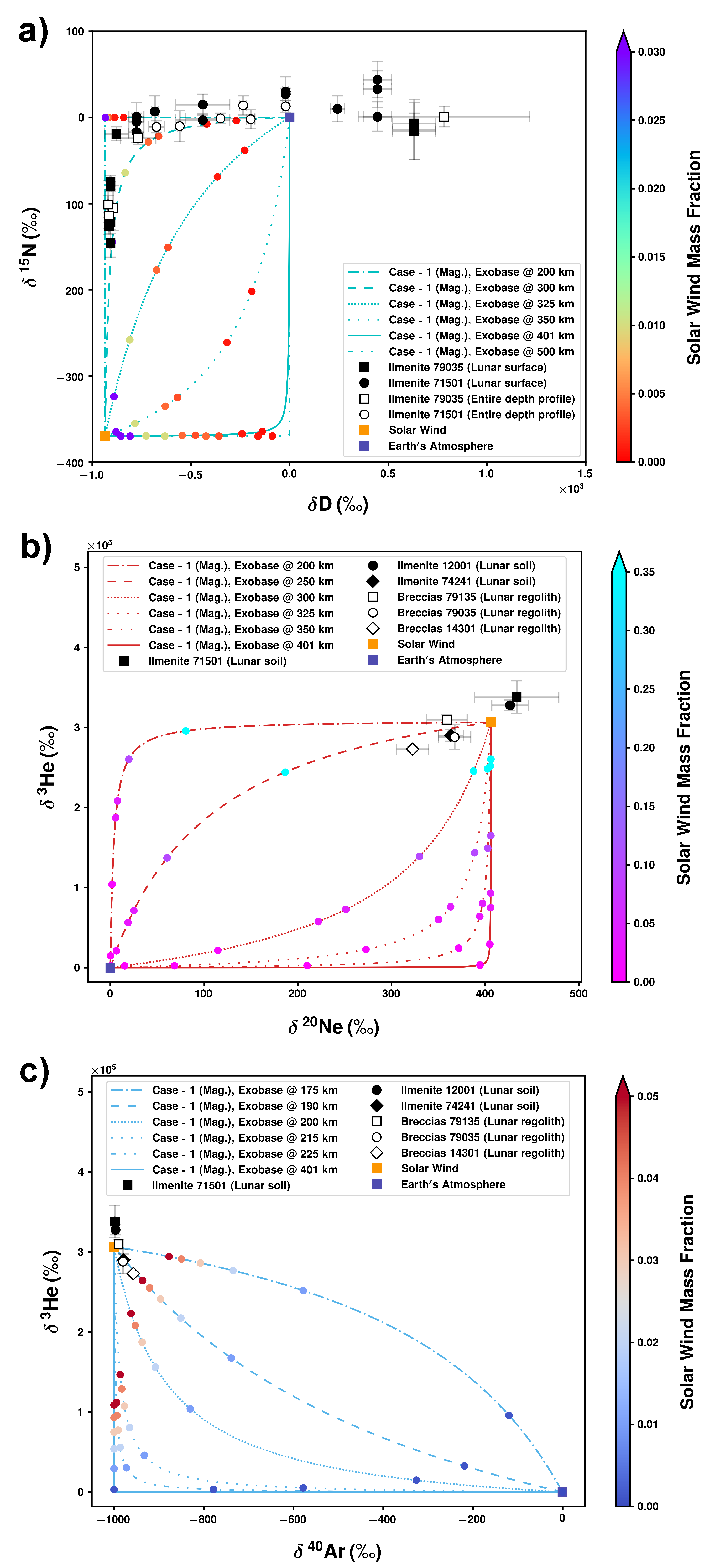}
    \caption{Description continued on the following page.}
    \label{fig:6}
\end{figure}

\begin{figure}
    \ContinuedFloat
    \caption{\textbf{Mixing diagrams for contemporary Earth with a core dynamo.} \textit{Top panel (a)} [$\mathrm{\delta\,{}^{15}N - \delta\, D}$ mixing diagram]: A mixing curve between the solar wind component (orange square) and the Earth wind component (navy square) in the magnetized case (Case -- \text{I}) is constructed with the estimated parameters that depend on the end member isotopic compositions and the curvature parameter ($K_{\mathsmaller{c}} = \mathrm{[ N / H ]_{\mathsmaller{\mathrm{SW}}}\,\big/ \, [ N / H ]_{\mathsmaller{\mathrm{EW}}}}$). The elemental ratios of the end members are inferred from fluxes given in Tables ED3 and ED4. The mixing hyperbolae are for different probable exobase heights for a magnetized Earth. The solid line indicates the profile for the current exobase at $\sim 401$ km. Observational data for lunar ilmenite samples from Apollo-17 breccias are plotted with squares and circles, respectively, along with their corresponding error bars. Solid symbols represent mean values surface grain taken from the depth profile at $< 100$ nm depth. Open symbols represent mean values of whole grains for the entire depth profile ($\sim 500$ nm). For solar wind isotope ratios, we used $\mathrm{{}^{15}N / {}^{14}N} = 2.32 \times 10^{-3}$ and $\mathrm{D / H} = 1.01 \times 10^{-5}$ following Hashizume et al. \cite{Hashizume2000}. Following Ozima et al. \cite{Ozima2005}, we used the present-day terrestrial atmosphere isotope ratios, $\mathrm{{}^{15}N / {}^{14}N} = 3.68 \times 10^{-3}$ and $\mathrm{D / H} = 1.56 \times 10^{-4}$, for the Earth wind composition. \textit{Middle panel (b)} [$\mathrm{\delta\,{}^{3}He - \delta\, {}^{20}Ne}$ mixing diagram]: Mixing curves were generated for different exobase altitudes with respective solar and planetary abundances estimated from Tables ED3 and ED4. Data are obtained for ilmenite grains from lunar soils (solid markers) and regolith breccias (open markers). For the solar wind composition (orange square), we used $\mathrm{{}^{3}He / {}^{4}He} = 4.26 \times 10^{-4}$ and $\mathrm{{}^{20}Ne / {}^{22}Ne} = 13.78$. For the Earth wind isotopic ratios (navy square), we used the commonly accepted present atmospheric values, $\mathrm{{}^{3}He / {}^{4}He} = 1.38 \times 10^{-6}$ and $\mathrm{{}^{20}Ne / {}^{22}Ne} = 9.8$ \cite{White2023}. \textit{Bottom panel (c)} [$\mathrm{\delta\,{}^{3}He - \delta\, {}^{40}Ar}$ mixing diagram]: Six mixing curves are plotted with different elemental abundance ratios but fixed isotopic ratios assumed for the end-members. The only variable to draw a unique mixing curve connecting the two end-members with fixed isotopic ratios is the elemental abundance ratio, $K_{\mathsmaller{c}} = \mathrm{[ He / Ar ]_{\mathsmaller{\mathrm{SW}}}\,\big/ \, [ He / Ar ]_{\mathsmaller{\mathrm{EW}}}}$. We chose the generally assumed values of $\mathrm{{}^{3}He / {}^{4}He} = 4.26 \times 10^{-4}$ and $\mathrm{{}^{40}Ar / {}^{36}Ar} = 3 \times 10^{-4}$ for the solar wind ratio (orange square). For the Earth wind components (navy square), we used $\mathrm{{}^{3}He / {}^{4}He} = 1.38 \times 10^{-6}$ and $\mathrm{{}^{40}Ar / {}^{36}Ar} = 296.16$ to construct the mixing curves. Colored dots in each subplot denote the corresponding increments in \textcursive{\slshape \footnotesize f\;}$_{\mathsmaller{SW}}$.}    
\end{figure}

\clearpage


\newgeometry{top=2cm, bottom=1.5cm, left=0.85cm, right=0.75cm}

\begin{table*}
\centering
\captionsetup{justification=centering}
\caption{\textbf{Orbit-averaged values of solar wind and Earth wind fluxes reaching the Moon.} Solar and Earth wind fluxes impinging on the lunar nearside regolith averaged over different lunar phases ($-\,\pi$ to $\pi$ rad) as a function of solar wind density, its velocity, and terrestrial atmosphere density. The variations in these parameters constitute the different cases (\text{I} to \text{V}). Case -- \text{I} represents the contemporary Sun-Earth-Moon system. In subcases where the geodynamo is present (referred to as ``Magnetized Earth''), the surface density of the atmosphere is determined by integrating the present-day atmospheric profile over one simulation scale height ($H_{\mathsmaller{M}}$). In each case, the total atmospheric mass remains constant across the subcases, regardless of whether Earth possessed or lacked an intrinsic dynamo.}
 \label{tab:tab1}
\begin{tabular*}{\textwidth}{@{\extracolsep{\fill}}c|l@{\hspace*{15pt}}c@{\hspace*{15pt}}c@{\hspace*{15pt}}c@{\hspace*{15pt}}c@{\hspace*{15pt}}c@{\hspace*{5pt}}c}

\hline\hline
\rotatebox[origin=c]{90}{\textbf{}} & \thead[l]{Unmagnetized / \\ Magnetized Earth} & \thead[c]{Solar Wind \\ Velocity $\left(\frac{km}{s}\right)$} & \thead[c]{Solar Wind No.\\ Density $\left(\frac{1}{cm^3}\right)$} & \thead{Atmosphere Base\\ Density $\left(\frac{kg}{m^3}\right)$} & \thead{Solar Wind \\ Flux $\left(\frac{1}{m^2\text{-}s}\right)$} & \thead{Planet Flux \\$\left(\frac{1}{m^2\text{-}s}\right)$} \\
\hline\\[-5pt]

\multirow{4}{*}{\rotatebox[origin=c]{90}{\textbf{Case -- \text{I}}}} \\[-5pt] & \makecell[l]{Magnetized Earth} & $375.0$ & $7.0$& $1.64 \times 10^{-2}$ & $6.58 \times 10^{8}$ &  $4.41 \times 10^{6}$ \\[5pt]
& \makecell[l]{Unmagnetized Earth} & $375.0$ & $80.0$ & $1.05 \times 10^{-1}$ & $9.27 \times 10^{9}$ & $4.88 \times 10^{5}$ \\[18pt]

\hline\\[-5pt]

\multirow{4}{*}{\rotatebox[origin=c]{90}{\textbf{Case -- \text{II}}}} \\[-5pt] & \makecell[l]{Magnetized Earth} & $375.0$ & $3.0 \times 10^{2}$ & $9.48 \times 10^{-9}$ & $2.86 \times 10^{10}$ & $2.41 \times 10^{8}$ \\[5pt]
& \makecell[l]{Unmagnetized Earth} & $375.0$ & $3.0 \times 10^{3}$ & $7.50 \times 10^{-8}$ & $3.33 \times 10^{11}$ & $3.10 \times 10^{9}$ \\[20pt]

\hline\\[-5pt]

\multirow{4}{*}{\rotatebox[origin=c]{90}{\textbf{Case -- \text{III}}}} \\[-5pt] & \makecell[l]{Magnetized Earth} & $375.0$ & $5.0 \times 10^{2}$ & $1.64 \times 10^{-2}$ & $4.83 \times 10^{10}$ & $4.71 \times 10^{6}$ \\[5pt]
& \makecell[l]{Unmagnetized Earth} & $375.0$ & $5.0 \times 10^{2}$ & $1.05 \times 10^{-1}$ & $5.79 \times 10^{10}$ & $6.41 \times 10^{7}$ \\[21pt]

\hline\\[-5pt]

\multirow{4}{*}{\rotatebox[origin=c]{90}{\textbf{Case -- \text{IV}}}} \\[-5pt] & \makecell[l]{Magnetized Earth} & $375.0$ & $7.0$ & $1.64 \times 10^{-7}$ & $6.51 \times 10^{8}$ & $4.75 \times 10^{4}$ \\[5pt]
& 
\\[20pt]

\hline\\[-5pt]

\multirow{4}{*}{\rotatebox[origin=c]{90}{\textbf{Case -- \text{V}}}} \\[-5pt] & 
\\[5pt]
& \makecell[l]{Unmagnetized Earth} & $500.0$ & $7.0 \times 10^{3}$ & $1.05 \times 10^{-1}$ & $1.06 \times 10^{12}$ & $1.22 \times 10^{10}$ \\[19pt]

\hline\hline
\end{tabular*}
\end{table*}

\restoregeometry
\baselineskip24pt

\clearpage


\newgeometry{top=0.8cm, bottom=1.5cm, left=1.0cm, right=1.0cm}

\begin{sidewaystable}
\centering
\captionsetup{justification=centering}
\caption{\textbf{Solar, terrestrial atmosphere, and non-solar fluxes in lunar nearside regolith for both present-day and primordial conditions.} Comparison of the implanted Earth wind (computed using our wind-atmosphere model) and non-solar fluxes (estimated from lunar soil data) for present-day magnetic Earth with that during the putative unmagnetized phase of early Earth. The non-solar flux values are calculated by multiplying the non-solar fraction with the solar wind flux. The combined MHD-ionization model explains the data when the EW flux exceeds the measured non-solar flux.}
\label{tab:tab2}
\begin{tabular*}{\textheight}{@{\extracolsep{\fill}}l@{\hspace*{15pt}}cccc@{\hspace*{15pt}}cccc}
\hline\hline
\thead[l]{\\ \\ \\ \vspace{-9pt}Element} & \multicolumn{4}{c}{\thead{Magnetized Earth (Case -- \text{I})}} & \multicolumn{4}{c}{\thead{Unmagnetized Earth (Case -- \text{I})}} \\
\cline{2-5} \cline{6-9}\addlinespace[3pt]
& \thead{Solar Wind \\ Flux $\left(\frac{1}{m^2\text{-}s}\right)$} & \thead{Non-Solar \\ Fraction} & \thead{Measured Non - \\ Solar Flux $\left(\frac{1}{m^2\text{-}s}\right)$} & \thead{Earth Wind \\ Flux $\left(\frac{1}{m^2\text{-}s}\right)$} & \thead{Solar Wind \\ Flux $\left(\frac{1}{m^2\text{-}s}\right)$} & \thead{Non-Solar \\ Fraction} & \thead{Measured Non - \\ Solar Flux $\left(\frac{1}{m^2\text{-}s}\right)$} & \thead{Earth Wind \\ Flux $\left(\frac{1}{m^2\text{-}s}\right)$} \\
\addlinespace[3pt]
\hline\\[5pt]

\makecell[l]{H}  & $6.31 \times 10^{8}$ & $0.193$ & $1.22 \times 10^{8}$ & $4.42 \times 10^{5}$ & $8.89 \times 10^{9}$ & $0.194$ & $1.72 \times 10^{9}$ & $4.72 \times 10^{5}$ \\[5pt]

\makecell[l]{He} & $2.71 \times 10^{7}$ & $0.015$ & $4.18 \times 10^{5}$ & $3.73 \times 10^{6}$ & $3.82 \times 10^{8}$ & $0.084$ & $3.21 \times 10^{7}$ & $1.98 \times 10^{4}$ \\[5pt]

\makecell[l]{N} & $4.37 \times 10^{4}$ & $0.449$ & $1.96 \times 10^{4}$ & $2.41 \times 10^{5}$ & $6.16 \times 10^{5}$ & $0.857$ & $5.28 \times 10^{5}$ & $3.48 \times 10^{3}$ \\[5pt]

\makecell[l]{Ne} & $1.51 \times 10^{3}$ & $0.025$ & $3.78 \times 10^{1}$ & $9.57 \times 10^{2}$ & $2.13 \times 10^{4}$ & $0.131$ & $2.78 \times 10^{3}$ & $1.92$ \\[5pt]

\makecell[l]{Ar} & $9.46 \times 10^{4}$ & $0.003$ & $2.46 \times 10^{2}$ & $3.38 \times 10^{2}$ & $1.33 \times 10^{6}$ & $0.047$ & $6.21 \times 10^{4}$ & $1.42 \times 10^{1}$ \\[20pt]
\hline\hline
\end{tabular*}
\end{sidewaystable}

\restoregeometry
\baselineskip24pt
\clearpage


\section*{Data and Materials Availability}
All data needed to evaluate the conclusions in the paper are present in the paper and/or the Supplementary Materials.

\section*{Code Availability}
The code \href{https://bluehound2.circ.rochester.edu/astrobear/}{\normalsize{AstroBEAR}} is freely available online\footnote[2]{\href{https://bluehound2.circ.rochester.edu/astrobear/}{\color{natgreen}http://astrobear.pas.rochester.edu/}}. The simulation outputs and Python routines are available upon request from the corresponding author.

\section*{Acknowledgments}
\commentout{We thank the manuscript reviewers and the handling editor for their valuable comments and suggestions that helped us enhance the quality of this work. }This work used high-performance computing resources and visualization tools provided by the Center for Integrated Research Computing (CIRC) at the University of Rochester (UofR). The authors thank CIRC for their technical support throughout this work. We also thank Colin Johnstone for providing the Archean homopause model data and for discussions. \textbf{Funding:} This work was supported by the Horton Graduate Fellowship from UofR's Laboratory for Laser Energetics (to S.P.), as well as NSF grant EAR2051550 and NASA grant 80NSSC19K0510 (to J.A.T.). E.G.B. acknowledges work at the Aspen Center for Physics, supported by NSF grant PHY-2210452, as well as additional support from NSF grant PHY-2020249 and DOE grant DE-SC0021990.

\section*{Author Contributions}
E.G.B. and J.A.T. contributed to the idea of the project; S.P. and E.G.B. jointly designed the project; S.P., J.C.-N., and E.G.B. conducted the MHD simulations; S.P. and E.G.B. performed research; S.P. wrote most of the initial draft of the paper, while E.G.B. reviewed and edited the manuscript. All authors contributed to the discussion, interpretation of the results, and revising the manuscript.

\section*{Competing Financial Interests}
The authors declare that they have no competing financial or non-financial interests.

\newpage



\newpage

\vspace*{16pt}

\begin{center}
{\fontsize{18}{22}\selectfont
Supplementary Materials for
}

\vspace*{16pt}

{\fontsize{14}{17}\selectfont
\textbf{Terrestrial atmospheric ion implantation occurred in the nearside lunar \\ regolith during the history of Earth's dynamo}
}

\vspace*{16pt}

{Shubhonkar Paramanick$^{\ddagger}$ \textit{et al.} 
}
\vspace*{13pt}

{\fontsize{10}{12}\selectfont
$^{\ddagger}$Corresponding author. Email: \email{shubhonkar.paramanick@rochester.edu}
}

\vspace*{0.7in}

\end{center}

\textbf{This PDF file includes:}
    \begin{description}[leftmargin=0em, labelindent=2em, itemsep=-0.1em]
        \item[$\square$] Supplementary Text
        \item[] \hspace{0.6em} $\circ$ \hspace{0.2em} Section S1: Limitations of the model
        \item[] \hspace{0.6em} $\circ$ \hspace{0.2em} Section S2: Analytic scaling relations governing atmospheric mass loss flux in the MHD regime
        \item[$\square$] Extended Data
        \item[] \hspace{0.6em} $\circ$ \hspace{0.2em} Figures ED1 to ED5
        \item[] \hspace{0.6em} $\circ$ \hspace{0.2em} Tables ED1 to ED5
    \end{description}

\newpage

\renewcommand{\figurename}{Extended Data Figure}
\renewcommand{\thefigure}{ED\arabic{figure}}
\setcounter{figure}{0}

\renewcommand{\tablename}{Extended Data Table}
\renewcommand{\thetable}{ED\arabic{table}}
\setcounter{table}{0}

\renewcommand{\thesection}{S\arabic{section}}
\renewcommand{\theHsection}{S\arabic{section}}
\renewcommand{\thesubsection}{\thesection.\arabic{subsection}}
\renewcommand{\thesubsubsection}{\thesubsection.\arabic{subsubsection}}
\setcounter{section}{0}

\setcounter{section}{0}


\begin{center}
{\fontsize{18}{22}\selectfont
\it{Supplementary Text}
}
\end{center}

\section{Analytic scaling relations governing atmospheric mass loss flux in the MHD regime}
We compare the density dependence of empirical scaling relations for atmospheric mass loss flux obtained from different MHD simulations with a simple fluid flow model for both magnetized and unmagnetized cases that offers some valuable insights into the complex plasma interaction model that was discussed earlier in this paper.

\subsection{Escaping mass flux from unmagnetized Earth}
Consider a SW flow  with velocity $\vec{v}_{\hspace{-0.2em}\mathsmaller{SW}}$ and  density and pressure  given by $\rho_{\mathsmaller{SW}}$ and $p_{\mathsmaller{SW}}$, respectively. This flow is incident upon Earth's atmosphere also modeled as a fluid. Let the atmospheric outflow velocity, mass density, and pressure be given by $\vec{v}_{\hspace{-0.2em}\mathsmaller{AO}}$, $\rho_{\mathsmaller{AO}}$, and $p_{\mathsmaller{AO}}$, respectively, in the quasi-steady state. Since momentum (and energy) of the fluid interaction must be conserved, we can readily relate the flow parameters on the dayside as follows: 
\begin{equation}
    p^{\left(\substack{\mathsmaller{Ram,} \\ \mathsmaller{SW}}\right)}\,+ \,p^{\left(\substack{\mathsmaller{Th,} \\ \mathsmaller{SW}}\right)} = \frac{\rho_{\mathsmaller{AO}}\,v_{\mathsmaller{AO}}^2}{2}\, + p_{\mathsmaller{AO}}
    \label{eq:eq20}
\end{equation}
The thermal pressure and magnetic pressure exerted by the SW are assumed negligible  compared to the SW ram pressure, as they are typically one to two orders of magnitude smaller. At the stagnation point, there is an exact balance between the SW ram pressure and the atmospheric thermal pressure, and the flow squeezes out from that point. Thus, the outflow speed is a manifestation of the thermal pressure. This results in the relation
\begin{equation}
    \frac{\rho_{\mathsmaller{SW}}\,v_{\mathsmaller{SW}}^2}{2} = \frac{\rho_{\mathsmaller{AO}}\,v_{\mathsmaller{AO}}^2}{2}
    \label{eq:eq21}
\end{equation}
Defining the outflow gas pressure in terms of the sound speed $c_{\mathsmaller{AO}} = \left[\gamma_{\mathsmaller{AO}}\, p_{\mathsmaller{AO}} / \rho_{\mathsmaller{AO}} \right]^{\frac{1}{2}} $ on the dayside, we obtain the expression for the escaping flux
\begin{equation}
    \mathcal{F}_{\mathsmaller{A},\,\mathsmaller{UM}}^{\mathsmaller{(Day)}}  = \rho_{\mathsmaller{AO}} \, v_{\mathsmaller{AO}} = \dfrac{\gamma_{\mathsmaller{AO}}\, p_{\mathsmaller{AO}}\, v_{\mathsmaller{AO}}}{c_{\mathsmaller{AO}}^{2}}
    \label{eq:eq22}
\end{equation}
The atmospheric mass flux escaping from the nightside at the lunar location in the unmagnetized case is given by
\begin{equation}
    \mathcal{F}_{\mathsmaller{A},\,\mathsmaller{UM}} = \zeta_{\mathsmaller{UM}} \sqrt{\rho_{\mathsmaller{AO}}\, \rho_{\mathsmaller{SW}}\, v_{\mathsmaller{SW}}^2}
    \label{eq:eq22a}
\end{equation}
The dimensionless constant $\zeta_{\mathsmaller{UM}}$, which is determined from MHD simulations, represents the decrease in planetary flux from near the plasmapause to the Moon and any difference in mass loading fraction between the dayside and nightside. For unmagnetized Earth, the SW impinges and ablates the top of the atmosphere. So, the above expression for mass flux can be rewritten as a function of atmosphere surface density as
\begin{equation}
    \mathcal{F}_{\mathsmaller{A},\,\mathsmaller{UM}} = \zeta_{\mathsmaller{UM}} \left[\rho_{\circ,{\mathsmaller{UM}}}\; \rho_{\mathsmaller{SW}}\, v_{\mathsmaller{SW}}^2\, e^{-\frac{z_{\mathsmaller{PP}}}{H_{\mathsmaller{UM}}}} \right]^{\frac{1}{2}} 
    \label{eq:eq23}
\end{equation}
This implies that the flux is proportional to the square root of product of the base density and SW density (Figure ED5).

\subsection{Escaping mass flux from magnetized Earth}

For the present Earth with an intrinsic dipolar field, the total pressure in magnetized plasma is given as the sum of the kinetic pressure, the thermal pressure, and the magnetic pressure. Since the SW is dominated by the dynamic pressure, and the flow frame pressure, away from the current sheet in the magnetotail is dominated by the magnetic rather than thermal pressure, the momentum conservation (\autoref{eq:eq20}) in the dayside region is modified to
\begin{equation}
    \frac{\rho_{\mathsmaller{SW}}\,v_{\mathsmaller{SW}}^2}{2} = \frac{\rho_{\mathsmaller{AO}}\,v_{\mathsmaller{AO}}^2}{2}\, + \frac{\rho_{\mathsmaller{AO}}\,v_{\mathsmaller{Alf},{\mathsmaller{AO}}}^2}{2}
    \label{eq:eq25}
\end{equation}
Here $v_{\mathsmaller{Alf},{\mathsmaller{AO}}} = \left[B_{\mathsmaller{GF}}^2 \,/ \,\mu_{\circ}\,\rho_{\mathsmaller{AO}} \right]^{\frac{1}{2}} $ is the Alfv$\Acute{\text{e}}$n velocity of the outflow. The nightside, where a significant amount of planetary material escapes, exhibits a low magnetic field strength. The escaping mass flux from magnetized Earth simplifies to
\begin{equation}
    \mathcal{F}_{\mathsmaller{A},\,\mathsmaller{M}} = \zeta_{\mathsmaller{M}} \left[\frac{\rho_{\circ,{\mathsmaller{M}}}\; e^{-\frac{r_{\mathsmaller{\beta_p}}-\;R_{\mathsmaller{\oplus}}}{H_{\mathsmaller{M}}}}}{\left[\frac{r_{\mathsmaller{MP}}}{r_{\mathsmaller{\beta_p}}}\right]^{\xi}}\; \rho_{\mathsmaller{SW}}\, v_{\mathsmaller{SW}}^2 \right]^{\frac{1}{2}}
    \label{eq:eq26}
\end{equation}
Thus, the mass flux again varies as the square root of product of the base density and SW density (Figure ED5).

Our semi-analytic scaling relations for the EW mass flux to the Moon depend only on the SW ram pressure, the terrestrial atmosphere surface density, the scale height, and a numerically determined geometric constant for both magnetized and unmagnetized Earth cases with different lunar orbits, corresponding to the contemporary and a primordial Earth-Moon system. The SW flux at the lunar location, on the other hand, exhibits an almost linear scaling with its ram pressure. These scaling relations are useful for estimating the SW and EW mass fluxes without having to extend the range of simulations.

\section{Limitations of the model}

\subsection{Planetary Rotation}

We do not include the effect of planet's rotation in our simulations. Given the Earth's relatively low angular velocity, we find that the axial components of both centrifugal and Coriolis forces are typically negligible compared to the magnetic force in the interaction region $\left(\mid 2\,\Vec{\Omega} \times \Vec{v} \mid \big/ \rho,\; \mid \Vec{\Omega} \times \Vec{\Omega} \times \Vec{r} \mid \big/ \rho \ll \frac{B^2}{2} \right)$. Hence, we do not benefit much from using a frame of reference co-rotating with the planet.

\subsection{Cartesian Grid and Diffusion}

AstroBEAR currently lacks the capability to conduct simulations in spherical coordinates, resulting in minor, stair-step-like artifacts within our planetary grid models. These artifacts tend to diminish as the resolution or mesh refinement level is increased, but the incremental gain in computational precision does not justify the additional time and resources required.

For simulations involving the present-day geomagnetic field, we observed certain cells on reconnection sites becoming over-saturated with magnetic field values, leading to a reduction in global time step size and eventual simulation slowdown. To address this issue, we implemented controlled numerical diffusivity within the simulation box, which facilitated quicker attainment of steady-state convergence and effectively mitigated the problem.

\clearpage

\begin{center}
{\fontsize{18}{22}\selectfont
\it{Extended Data}
}
\end{center}


\begin{figure}[H]
    \centering
	\includegraphics[scale=0.43]{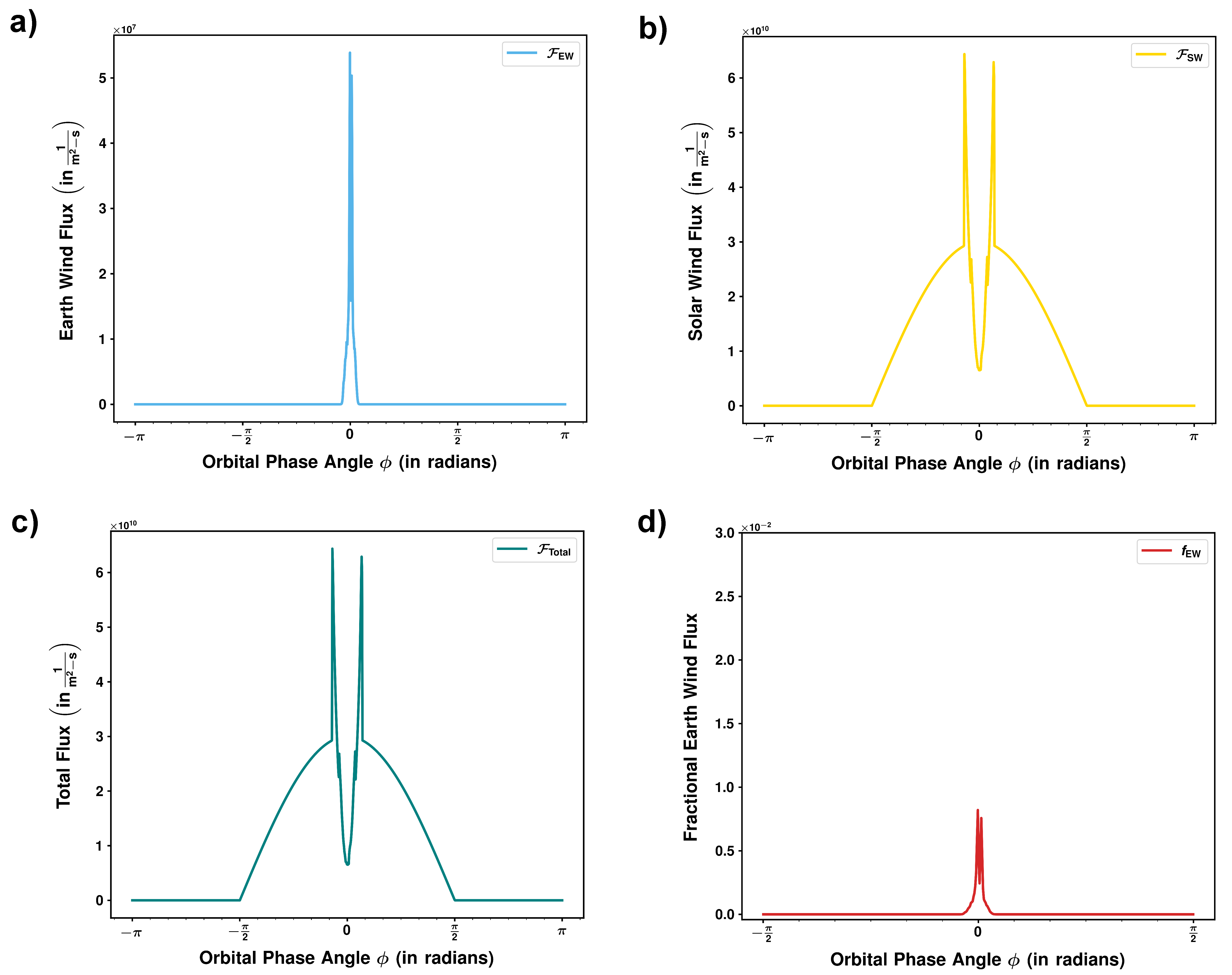}
    \caption{\textbf{Orbital variations of species-total mass fluxes impinging on the lunar surface in the unmagnetized Earth case.} Angular distribution of the \textit{(a)} planet atmosphere flux, \textit{(b)} solar wind flux, \textit{(c)} total flux (aggregate flux of the terrestrial atmosphere and solar wind), and \textit{(d)} fractional contribution of planetary wind to the total flux over a full lunation. At phase angles $-\pi,\, \pi$ rad, the Moon is in the dayside upstream solar wind at the sub-solar line of Earth. The double-horned solar wind and total flux profiles result from shocked material located between the magnetosheath and the magnetotail lobe.}
    \label{fig:ED1}
\end{figure}

\clearpage



\begin{figure}
    \centering
	\includegraphics[scale=0.45]{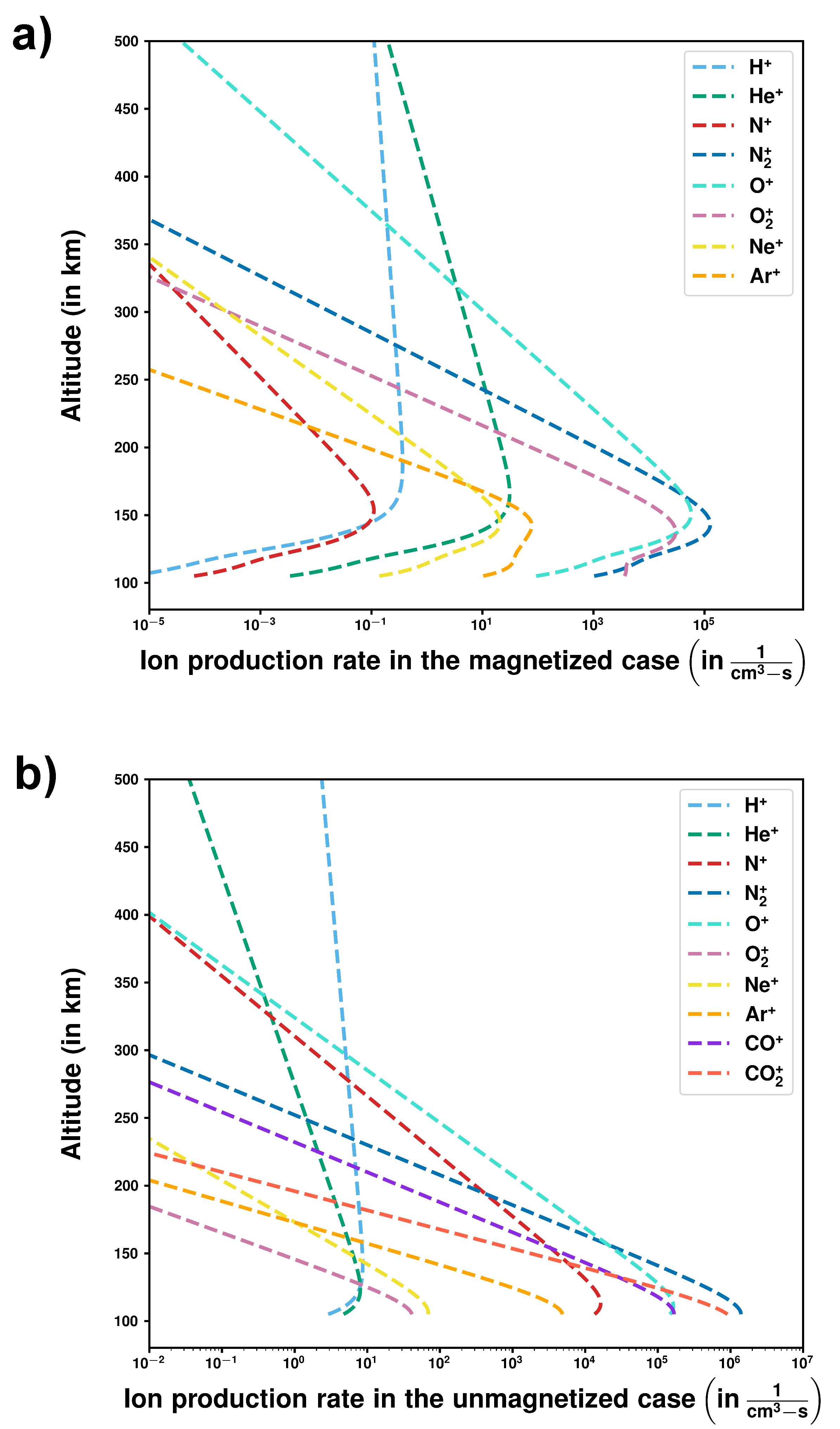}
    \caption{\textbf{Altitude profiles of the production rates of volatile ions by photoionization for present-day Earth (with a geodynamo) and Archean Earth (lacking a geodynamo).} \textit{Top panel (a)}: Ionization rates of H$^{+}$ (sky blue), He$^{+}$ (forest green), N$^{+}$ (crimson), N$_{2}^{+}$ (cerulean blue), O$^{+}$ (turquoise), O$_{2}^{+}$ (sky magenta), Ne$^{+}$ (dandelion yellow), and Ar$^{+}$ (orange) ions estimated from a semi-theoretical photoionization model (EUVAC) and an empirical model (NRLMISE-00) of the neutral densities of the constituents in the terrestrial atmosphere under normal solar EUV flux condition. \textit{Bottom panel (b)}: Similar to \textit{panel (a)}, but also includes CO$^{+}$ (lavender) and CO$_{2}^{+}$ (coral) for the unmagnetized Earth with Archean terrestrial upper atmosphere, representing primordial Earth case.}
    \label{fig:ED2}
\end{figure}

\clearpage



\begin{figure}
    \centering
	\includegraphics[scale=0.43]{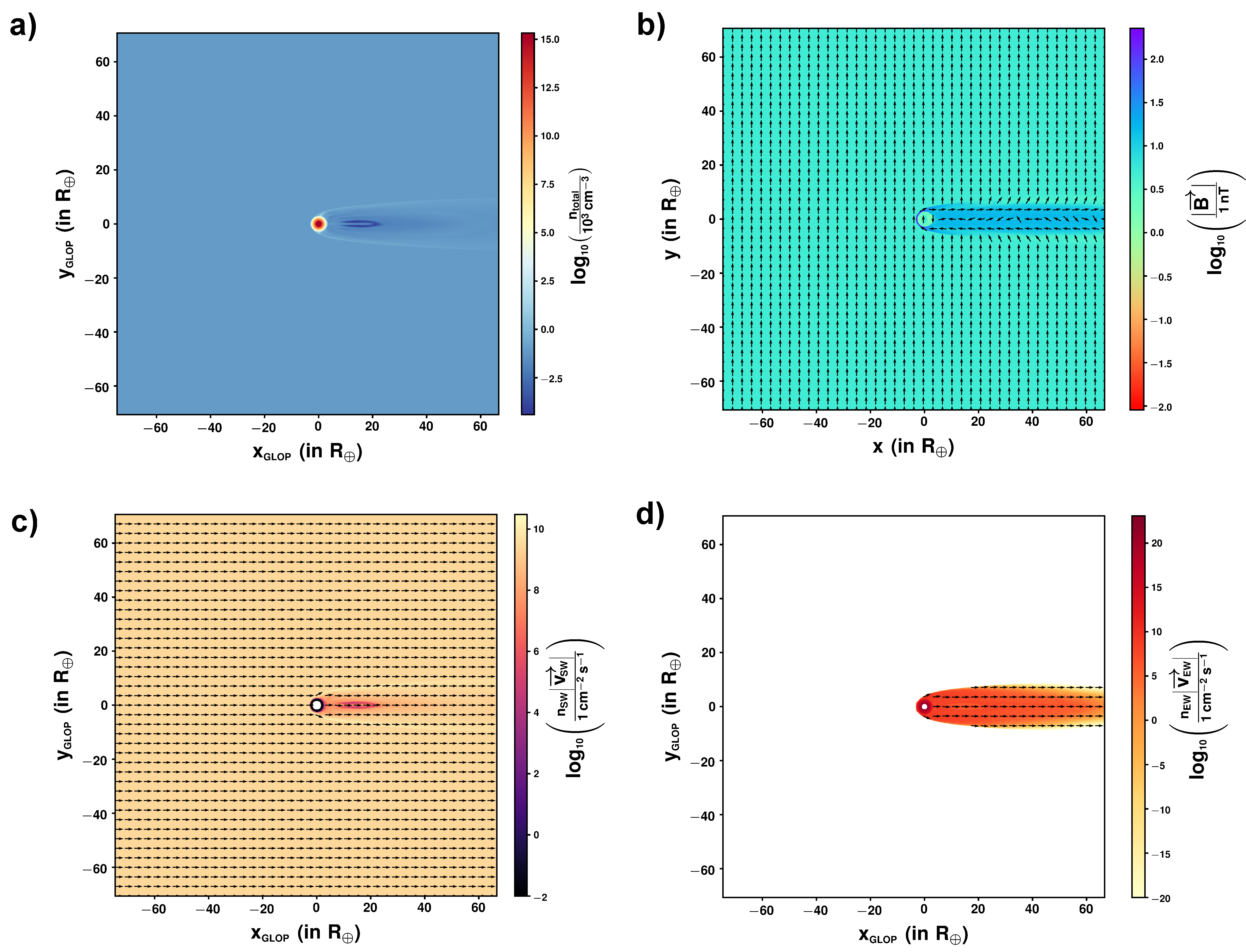}
    \caption{\textbf{Results from our 3D-MHD simulation model for the case in which the planet lacks an intrinsic dynamo (Case -- \text{I}).} \textit{Panel (a)}: Total number density plot depicting the large-scale structure of the wind-atmosphere interaction upto the lunar orbit in the quasi-steady state. \textit{Panel (b)}: Polar-cut view of the topology of the planet's induced magnetosphere under the influence of primordial magnetized solar wind. Quivers represent the induced magnetic field. \textit{Panel (c)}: 2D slice of the normalized solar wind mass flux in the lunar orbital plane. The flux drops by $\sim 3$ orders of magnitude inside the magnetotail compared to the shocked solar wind region between the bow shock and the plasmapause boundary, where it exhibits a gradual gradient. \textit{Panel (d)}: Terrestrial atmosphere mass flux in the magnetotail region. The quiver arrows denote the plane-projected bulk velocity component indicating the escaping pathways. The filled white circle represents the planet's core, which is not included in the passive advective tracers.}
    \label{fig:ED3}
\end{figure}

\clearpage


\begin{figure}
    \centering
	\includegraphics[scale=0.45]{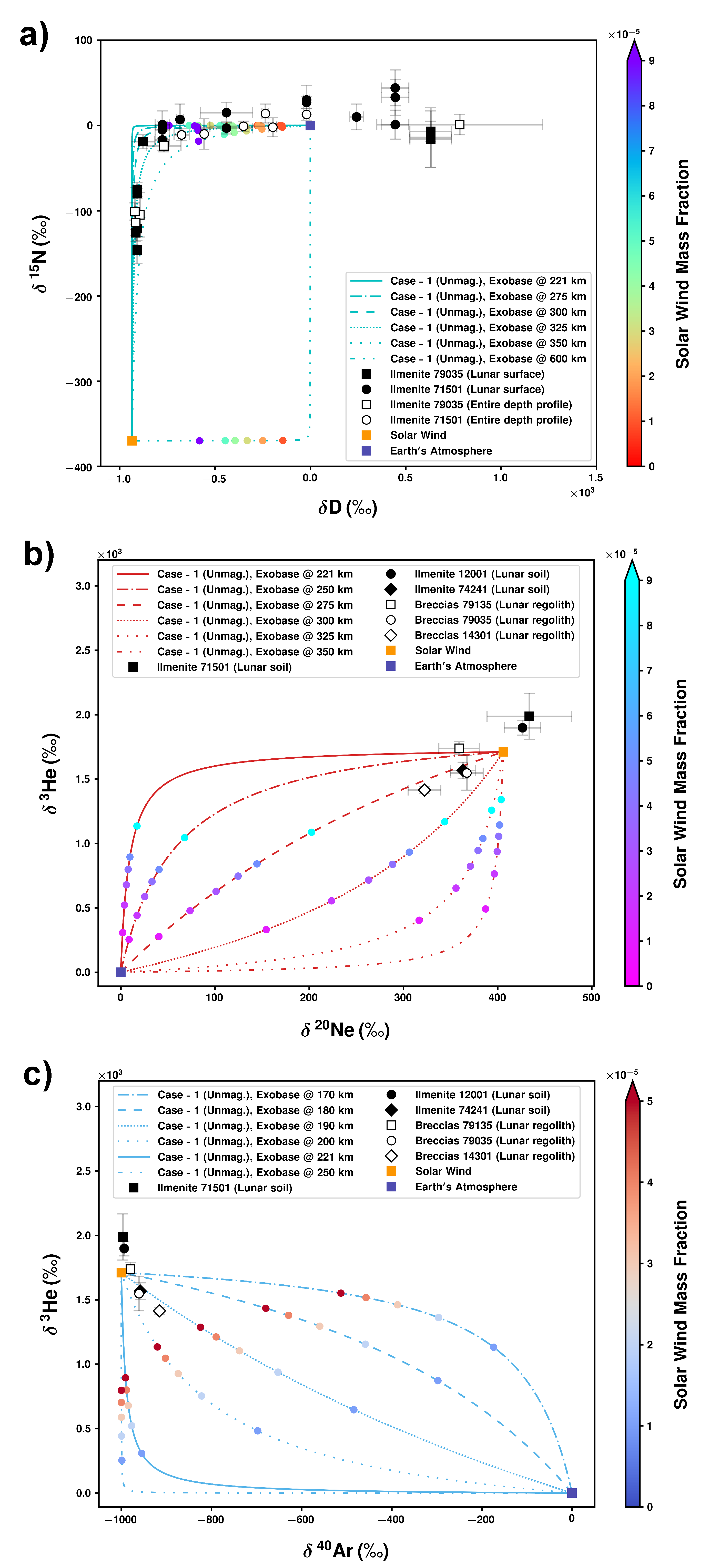}
    \caption{Description continued on the following page.}
    \label{fig:ED4}
\end{figure}

\begin{figure}
    \ContinuedFloat
    \caption{\textbf{Mixing diagrams for Archean Earth without an intrinsic dipole field.} \textit{Top panel (a)} [$\mathrm{\delta\,{}^{15}N - \delta\, D}$ mixing diagram]: A mixing curve between the solar wind component (orange square) and the Earth wind component (navy square) in the unmagnetized case (Case -- \text{I}) representing Eoarchean eon is constructed with the estimated parameters. The elemental ratios of the end members are inferred from fluxes given in Tables ED3 and ED5. The mixing hyperbolae are for different probable exobase heights for a non-magnetic Earth. The solid line indicates the profile for the primitive Earth's exobase at $\sim 221$ km. Observational data for lunar ilmenite samples from Apollo-17 breccias are plotted with squares and circles, respectively, along with their corresponding error bars. For solar wind isotope ratios, we used $\mathrm{{}^{15}N / {}^{14}N} = 2.32 \times 10^{-3}$ and $\mathrm{D / H} = 1.01 \times 10^{-5}$. We used the Archean epoch terrestrial atmosphere isotope ratios, $\mathrm{{}^{15}N / {}^{14}N} = 3.68 \times 10^{-3}$ and $\mathrm{D / H} = 1.56 \times 10^{-4}$, for the Earth wind composition. \textit{Middle panel (b)} [$\mathrm{\delta\,{}^{3}He - \delta\, {}^{20}Ne}$ mixing diagram]: Mixing curves were generated for different exobase altitudes with respective solar and planetary abundances estimated from Tables ED3 and ED5. Data are obtained for ilmenite grains from lunar soils (solid markers) and regolith breccias (open markers). For the solar wind composition (orange square), we used $\mathrm{{}^{3}He / {}^{4}He} = 4.26 \times 10^{-4}$ and $\mathrm{{}^{20}Ne / {}^{22}Ne} = 13.78$. For the Earth wind isotopic ratios (navy square), we used the commonly accepted value of $\mathrm{{}^{20}Ne / {}^{22}Ne} = 9.8$ and primordial terrestrial ratio of $\mathrm{{}^{3}He / {}^{4}He} = 1.57 \times 10^{-4}$ following Ozima et al. \cite{Ozima2005}. \textit{Bottom panel (c)} [$\mathrm{\delta\,{}^{3}He - \delta\, {}^{40}Ar}$ mixing diagram]: Six mixing curves are plotted with different elemental abundance ratios but fixed isotopic ratios assumed for the end-members. We chose the generally assumed values of $\mathrm{{}^{3}He / {}^{4}He} = 4.26 \times 10^{-4}$ and $\mathrm{{}^{40}Ar / {}^{36}Ar} = 3 \times 10^{-4}$ for the solar wind ratio (orange square). For the ancient Earth wind components (navy square), we used $\mathrm{{}^{3}He / {}^{4}He} = 1.57 \times 10^{-4}$ and $\mathrm{{}^{40}Ar / {}^{36}Ar} = 150.0$ to construct the mixing curves. Colored dots in each subplot illustrate the corresponding increments in \textcursive{\slshape \footnotesize f\;}$_{\mathsmaller{SW}}$.
    }
\end{figure}

\clearpage



\begin{figure}
    \centering
	\includegraphics[scale=0.5]{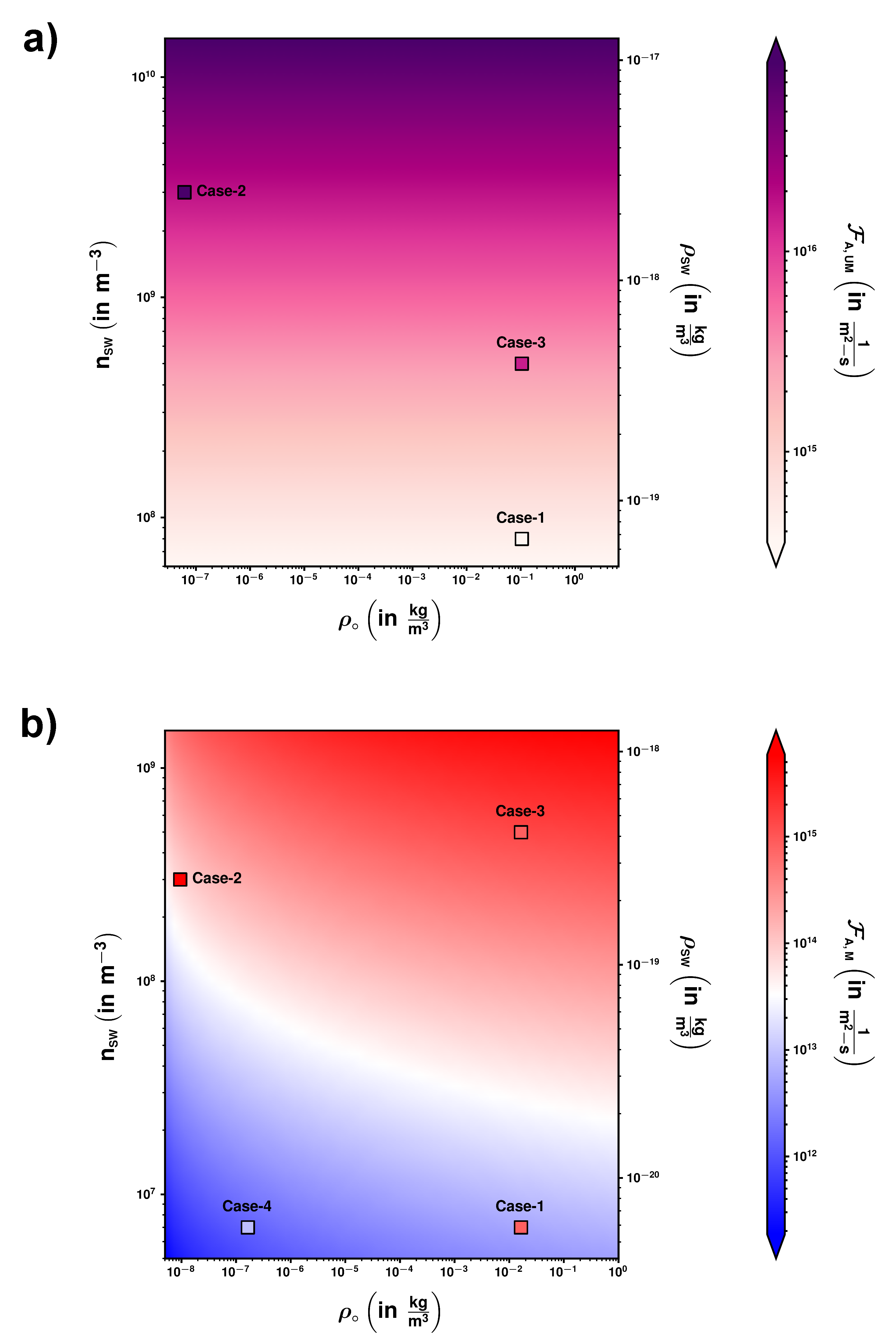}

    \caption{\textbf{Correlation of the planetary flux scaling laws $\left(\mathcal{F}_{\mathsmaller{A},\,\mathsmaller{UM}} \;\&\; \mathcal{F}_{\mathsmaller{A},\,\mathsmaller{M}}\right)$ with the terrestrial atmosphere surface density ($\rho_{\circ}$) and solar wind density $\left(n_{\mathsmaller{SW}}\right)$ in \textit{(a)} the absence and \textit{(b)} the presence of a geodynamo.} Overlaid on the 2-D colormaps of the scaling relations are the planetary flux values (represented with squares) computed from the 3-D MHD simulations (refer to \autoref{tab:tab1}). The abscissa limits for the unmagnetized case are determined by maintaining the total atmospheric mass constant. The simulation datapoints best follow the analytic expressions in \hyperref[eq:eq23]{Equations 40} and \ref{eq:eq26} if their color matches that of the background.}
    \label{fig:ED5}
\end{figure}

\clearpage


\newgeometry{top=0.8cm, bottom=1.5cm, left=1.0cm, right=1.0cm}

\begin{table}
\centering
\begin{threeparttable}
\captionsetup{justification=centering}
\caption{\textbf{Stellar wind and planetary atmosphere parameters used in the simulations for the magnetized cases.} The ambient medium here refers to the region outside the magnetopause boundary at the start of the simulation, where we adopt the same values as those of the solar wind. Across the various simulation cases, most parameters are kept fixed, except for those that directly influence atmospheric escape, to examine their effects.}
 \label{tab:tabED1}
 \begin{tabular}{lccc}

  \hline\hline 
  \thead[l]{Stellar\\parameters} & \thead{Value used} & \thead{Notation} & \thead{Units}\\
  \hline\\[-5pt]  
  \makecell[l]{Radius} & $696.340$ &  $R_{\mathsmaller{\odot}}$ & $\mathrm{Mm}$\\[5pt] 
  \makecell[l]{Mass} & $1.99 \times 10^{30}$ &  $M_{\mathsmaller{\odot}}$ & $\mathrm{kg}$\\[5pt]   
 \makecell[l]{Wind density} & \makecell{Refer to \\ \autoref{tab:tab1}} &  $\rho_{\mathsmaller{SW}}$ & $\dfrac{\text{1}}{\text{cm}^3}$\\[5pt] 
  \makecell[l]{Wind\\temperature} & $1.0 \times 10^{5}$ &  $T_{\mathsmaller{SW}}$ & $\mathrm{K}$\\[10pt]  
  \makecell[l]{Wind\\ magnetic field} & $5.0$ & $B_{\mathsmaller{SW}}$ & $\mathrm{nT}$\\[10pt]
  
  \makecell[l]{Wind velocity} & \makecell{Refer to \\ \autoref{tab:tab1}} &  $v_{\mathsmaller{SW}}$ & $\dfrac{\text{km}}{\text{s}}$\\[10pt] 
  
  \hline\hline 
  \thead[l]{Planetary\\parameters} & \thead{Value used} & \thead{Notation} & \thead{Unit}\\
  \hline\\[-5pt]

  \makecell[l]{Radius} & $6.378$ &  $R_{\mathsmaller{\oplus}}$ & $\mathrm{Mm}$\\[5pt]   
  \makecell[l]{Core density} & $5.51 \times 10^{3}$ &  $\rho_{\mathsmaller{\oplus}}$ & $\dfrac{\text{kg}}{\text{m}^3}$ \\[9pt]  
  \makecell[l]{Atmosphere \\surface density } & \makecell{Refer to \\ \autoref{tab:tab1}} & $\rho_{\circ,\,{\mathsmaller{M}}}$ & $\dfrac{\text{kg}}{\text{m}^3}$ \\[10pt]  
 \makecell[l]{Atmospheric\\temperature} & $298.0$ & $T_{\circ}$ & $\mathrm{K}$\\[10pt] 
  \makecell[l]{Adiabatic index} & $1.0000001$ & $\gamma$ & $\text{---}$\\[5pt]  
  \makecell[l]{Planetary eq.\\magnetic field } & $3.12 \times 10^{4}$ & $B_{\mathsmaller{\oplus}}$ & $\mathrm{nT}$\\[10pt]  
  \makecell[l]{Magnetospheric\\tilt angles} & $0^{\circ},\; 11^{\circ},\; 90^{\circ}$ & $\psi$ & $\text{---}$ \\[10pt]
  
  \hline\hline 
  \thead[l]{Ambient\\parameters} & \thead{Value used} & \thead{Notation} & \thead{Unit}\\
  \hline\\[-5pt]

 \makecell[l]{Ambient density} & \makecell{Same as SW \\density} &  $\rho_{\mathsmaller{Amb}}$ & $\dfrac{\text{1}}{\text{cm}^3}$\\[5pt] 
  \makecell[l]{Ambient\\temperature} & $1.0 \times 10^{5}$ &  $T_{\mathsmaller{Amb}}$ & $\mathrm{K}$\\[10pt]  
  \makecell[l]{Ambient\\ magnetic field} & $5.0$ & $B_{\mathsmaller{Amb}}$ & $\mathrm{nT}$\\[10pt]  
  \makecell[l]{Ambient velocity} & \makecell{Same as SW \\velocity} &  $v_{\mathsmaller{Amb}}$ & $\dfrac{\text{km}}{\text{s}}$\\[10pt] 
  
  \hline\hline
  \end{tabular}
\end{threeparttable}
\end{table}

\restoregeometry
\baselineskip24pt
\clearpage



\begin{table*}
\centering
\begin{threeparttable}
\captionsetup{justification=centering}
\caption{\textbf{Stellar wind and planetary atmosphere parameters used in the simulations for the unmagnetized cases.\protect\tnotex{tn:1}} The ambient medium here refers to the region outside the plasmapause boundary at the start of the simulation, where we adopt the same values as those of the solar wind. Across the different simulation cases, most parameters remain constant except for those directly affecting atmospheric escape, in order to study their effect.}
 \label{tab:tabED2}

 \begin{tabular}{lccc}

  \hline\hline 
  \thead[l]{Stellar\\parameters} & \thead{Value used} & \thead{Notation} & \thead{Units}\\
  \hline\\[-5pt]

 \makecell[l]{Wind density} & \makecell{Refer to \\ \autoref{tab:tab1}} &  $\rho_{\mathsmaller{SW}}$ & $\dfrac{\text{1}}{\text{cm}^3}$\\[5pt]
  \makecell[l]{Wind\\ magnetic field} & $10.0$ & $B_{\mathsmaller{SW}}$ & $\mathrm{nT}$\\[10pt]
  \makecell[l]{Wind velocity} & \makecell{Refer to \\ \autoref{tab:tab1}} &  $v_{\mathsmaller{SW}}$ & $\dfrac{\text{km}}{\text{s}}$\\[10pt] 
  
  \hline\hline 
  \thead[l]{Planetary\\parameters} & \thead{Value used} & \thead{Notation} & \thead{Unit}\\
  \hline\\[-5pt]

  \makecell[l]{Atmosphere \\surface density } & \makecell{Refer to \\ \autoref{tab:tab1}} \commentout{$1.049\times 10^{-1}$} & $\rho_{\circ,\,{\mathsmaller{UM}}}$ & $\dfrac{\text{kg}}{\text{m}^3}$ \\[10pt]

  \hline\hline 
  \thead[l]{Ambient\\parameters} & \thead{Value used} & \thead{Notation} & \thead{Unit}\\
  \hline\\[-5pt]

 \makecell[l]{Ambient density} & \makecell{Same as SW \\density} &  $\rho_{\mathsmaller{Amb}}$ & $\dfrac{\text{1}}{\text{cm}^3}$\\[5pt]
  \makecell[l]{Ambient\\ magnetic field} & $10.0$ & $B_{\mathsmaller{Amb}}$ & $\mathrm{nT}$\\[10pt]
  \makecell[l]{Ambient velocity} & \makecell{Same as SW \\velocity} &  $v_{\mathsmaller{Amb}}$ & $\dfrac{\text{km}}{\text{s}}$\\[10pt] 
  
  \hline\hline
\end{tabular}
\begin{tablenotes}
 \item[\textdagger] \label{tn:1} The physical parameters of the star-planet system, along with the values used to initialize the winds during the period when the Earth lacked an intrinsic magnetic field, are outlined in this table. The remaining parameters are set to the same values as those presented in Table ED1.
\end{tablenotes} 
\end{threeparttable}
\end{table*}

\clearpage



\newgeometry{top=0.8cm, bottom=1.5cm, left=1.0cm, right=1.0cm}

\begin{sidewaystable}
\centering
\captionsetup{justification=centering}
\caption{\textbf{Relative contribution of each species to the solar wind flux reaching the Moon.} The relative elemental abundance of the solar wind is estimated from the observation data of the Ulysses and SOHO spacecraft \cite{Wiens2004}. The solar wind flux is then computed by multiplying this fraction with the orbit-averaged flux obtained from 3D-MHD simulations (Table 1).}
\label{tab:tabED3}
\begin{tabular*}{\textheight}{@{\extracolsep{\fill}}l@{\hspace*{15pt}}c@{\hspace*{15pt}}c@{\hspace*{5pt}}cc@{\hspace*{15pt}}cc@{\hspace*{15pt}}cc@{\hspace*{15pt}}cc@{\hspace*{15pt}}cc}
\hline\hline
\thead[l]{\\ \\ Element} & \thead{\\ \\ Relative \\ Fraction} & \multicolumn{2}{c}{\thead{Solar Wind Flux \\ in Case -- \text{I} $\left(\frac{1}{m^2\text{-}s}\right)$}} & \multicolumn{2}{c}{\thead{Solar Wind Flux \\ in Case -- \text{II} $\left(\frac{1}{m^2\text{-}s}\right)$}} & \multicolumn{2}{c}{\thead{Solar Wind Flux \\ in Case -- \text{III} $\left(\frac{1}{m^2\text{-}s}\right)$}} & \multicolumn{2}{c}{\thead{Solar Wind Flux \\ in Case -- \text{IV} $\left(\frac{1}{m^2\text{-}s}\right)$}} & \multicolumn{2}{c}{\thead{Solar Wind Flux \\ in Case -- \text{V} $\left(\frac{1}{m^2\text{-}s}\right)$}} \\
\cline{3-4} \cline{5-6} \cline{7-8} \cline{9-10} \cline{11-12}
& & \thead{Mag.\\Earth} 
& 
\thead{Unmag.\\Earth} 
& 
\thead{Mag.\\Earth} 
& 
\thead{Unmag.\\Earth} 
& 
\thead{Mag.\\Earth} 
& 
\thead{Unmag.\\Earth} 
& 
\thead{Mag.\\Earth} 
& 
& 
& 
\thead{Unmag.\\Earth} 
\\
\hline\\[5pt]

\makecell[l]{H}  & $9.59 \times 10^{-1}$ & $6.31 \times 10^{8}$ & $8.89 \times 10^{9}$ & $2.74 \times 10^{10}$ & $3.20 \times 10^{11}$ & $4.63 \times 10^{10}$ & $5.55 \times 10^{10}$ & $6.24 \times 10^{8}$ &  &  & $1.01 \times 10^{12}$\\[5pt]

\makecell[l]{He} & $4.12 \times 10^{-2}$ & $2.71 \times 10^{7}$ & $3.82 \times 10^{8}$ & $1.18 \times 10^{9}$ & $1.37 \times 10^{10}$ & $1.99 \times 10^{9}$ & $2.39 \times 10^{9}$ & $2.68 \times 10^{7}$ &  &  & $4.35 \times 10^{10}$ \\[5pt]

\makecell[l]{N} & $6.65 \times 10^{-5}$ & $4.37 \times 10^{4}$ & $6.16 \times 10^{5}$ & $1.90 \times 10^{6}$ & $2.22 \times 10^{7}$ & $3.21 \times 10^{6}$ & $3.85 \times 10^{6}$ & $4.33 \times 10^{4}$ &  &  & $7.02 \times 10^{7}$ \\[5pt]

\makecell[l]{Ne} & $2.29 \times 10^{-6}$ & $1.51 \times 10^{3}$ & $2.13 \times 10^{4}$ & $6.55 \times 10^{4}$ & $7.64 \times 10^{5}$ & $1.11 \times 10^{5}$ & $1.33 \times 10^{5}$ & $1.49 \times 10^{3}$ &  &  & $2.42 \times 10^{6}$ \\[5pt]

\makecell[l]{Ar} & $1.44 \times 10^{-4}$ & $9.46 \times 10^{4}$ & $1.33 \times 10^{6}$ & $4.11 \times 10^{6}$ & $4.79 \times 10^{7}$ & $6.95 \times 10^{6}$ & $8.32 \times 10^{6}$ & $9.36 \times 10^{4}$ &  &  & $1.52 \times 10^{8}$ \\[20pt]

\hline\hline

\end{tabular*}
\end{sidewaystable}

\restoregeometry

\clearpage



\begin{sidewaystable}
\centering
\captionsetup{justification=centering}
\caption{\textbf{Relative contribution of each volatile species to the planet wind flux reaching the Moon in the magnetized case.} The relative elemental abundance of the Earth wind is computed for different exobase heights using the ion fluxes escaping from the present-day terrestrial atmosphere (Subfigure 4 (a)). The Earth wind flux is then calculated by multiplying this fraction with the orbit-averaged planetary flux obtained from 3D-MHD simulations (Table 1). Only the values for an exobase height of 401 km are tabulated here.}
\label{tab:tabED4}
\begin{tabular*}{\textheight}{@{\extracolsep{\fill}}l@{\hspace*{15pt}}c@{\hspace*{15pt}}c@{\hspace*{5pt}}cc@{\hspace*{15pt}}cc@{\hspace*{15pt}}cc@{\hspace*{15pt}}cc@{\hspace*{15pt}}cc}
\hline\hline\addlinespace[5pt]
\thead[l]{ Element} & \thead{Relative \\ Fraction} & \multicolumn{1}{c}{\thead{Atmos. Wind Flux \\ from Mag. Earth \\ in Case -- \text{I} $\left(\frac{1}{m^2\text{-}s}\right)$}} & \multicolumn{1}{c}{\thead{Atmos. Wind Flux \\ from Mag. Earth \\ in Case -- \text{II} $\left(\frac{1}{m^2\text{-}s}\right)$}} & \multicolumn{1}{c}{\thead{Atmos. Wind Flux \\ from Mag. Earth \\ in Case -- \text{III} $\left(\frac{1}{m^2\text{-}s}\right)$}} & \multicolumn{1}{c}{\thead{Atmos. Wind Flux \\ from Mag. Earth \\ in Case -- \text{IV} $\left(\frac{1}{m^2\text{-}s}\right)$}} 
\\

\addlinespace[5pt]\hline\\[5pt]

\makecell[l]{H}  & $4.12 \times 10^{-1}$ & $1.82 \times 10^{6}$ & $9.95 \times 10^{7}$ & $1.94 \times 10^{6}$ & $1.96 \times 10^{4}$ 
\\[5pt]

\makecell[l]{He} & $5.88 \times 10^{-1}$ & $2.59 \times 10^{6}$ & $1.42 \times 10^{8}$ & $2.77 \times 10^{6}$ & $2.79 \times 10^{4}$ 
\\[5pt]

\makecell[l]{N} & $7.23 \times 10^{-8}$ & $3.19 \times 10^{-1}$ & $1.74 \times 10^{1}$ & $3.41 \times 10^{-1}$ & $3.43 \times 10^{-3}$ 
\\[5pt]

\makecell[l]{Ne} & $1.02 \times 10^{-8}$ & $4.48 \times 10^{-2}$ & $2.45$ & $4.78 \times 10^{-2}$ & $4.82 \times 10^{-4}$ 
\\[5pt]

\makecell[l]{Ar} & $9.98 \times 10^{-17}$ & $4.40 \times 10^{-10}$ & $2.41 \times 10^{-8}$ & $4.70 \times 10^{-10}$ & $4.74 \times 10^{-12}$ 
\\[20pt]

\hline\hline

\end{tabular*}
\end{sidewaystable}

\clearpage



\begin{sidewaystable}
\centering
\captionsetup{justification=centering}
\caption{\textbf{Relative contribution of each volatile species to the planet wind flux reaching the Moon in the unmagnetized case.} The relative elemental abundance of the Earth wind is computed for different exobase heights using the ion fluxes escaping from the Eoarchean epoch atmosphere (Subfigure 4 (b)). The Earth wind flux is then calculated by multiplying this fraction with the orbit-averaged planetary flux obtained from 3D-MHD simulations (Table 1). Only the values for an exobase height of 221 km are tabulated here.}
\label{tab:tabED5}
\begin{tabular*}{\textheight}{@{\extracolsep{\fill}}l@{\hspace*{15pt}}c@{\hspace*{15pt}}c@{\hspace*{5pt}}cc@{\hspace*{15pt}}cc@{\hspace*{15pt}}cc@{\hspace*{15pt}}cc@{\hspace*{15pt}}cc}
\hline\hline\addlinespace[5pt]
\thead[l]{ Element} & \thead{Relative \\ Fraction} 
& \multicolumn{1}{c}{\thead{Atmos. Wind Flux \\ from Unmag. Earth \\ in Case -- \text{I} $\left(\frac{1}{m^2\text{-}s}\right)$}} 
& \multicolumn{1}{c}{\thead{Atmos. Wind Flux \\ from Unmag. Earth \\ in Case -- \text{II} $\left(\frac{1}{m^2\text{-}s}\right)$}} 
& \multicolumn{1}{c}{\thead{Atmos. Wind Flux \\ from Unmag. Earth \\ in Case -- \text{III} $\left(\frac{1}{m^2\text{-}s}\right)$}} 
& \multicolumn{1}{c}{\thead{Atmos. Wind Flux \\ from Unmag. Earth \\ in Case -- \text{IV} $\left(\frac{1}{m^2\text{-}s}\right)$}} 
\\

\addlinespace[5pt]\hline\\[5pt]

\makecell[l]{H}  & $4.29 \times 10^{-1}$ & $2.09 \times 10^{5}$ & $1.33 \times 10^{9}$ & $2.75 \times 10^{7}$ & $5.24 \times 10^{9}$  
\\[5pt]

\makecell[l]{He} & $3.57 \times 10^{-2}$ & $1.74 \times 10^{4}$ & $1.11 \times 10^{8}$ & $2.29 \times 10^{6}$ & $4.37 \times 10^{8}$  
\\[5pt]

\makecell[l]{N} & $5.35 \times 10^{-1}$ & $2.61 \times 10^{5}$ & $1.66 \times 10^{9}$ & $3.43 \times 10^{7}$ & $6.53 \times 10^{9}$  
\\[5pt]

\makecell[l]{Ne} & $8.88 \times 10^{-5}$ & $4.33 \times 10^{1}$ & $2.75 \times 10^{5}$ & $5.69 \times 10^{3}$ & $1.08 \times 10^{6}$  
\\[5pt]

\makecell[l]{Ar} & $1.29 \times 10^{-6}$ & $6.32 \times 10^{-1}$ & $4.02 \times 10^{3}$ & $8.30 \times 10^{1}$ & $1.58 \times 10^{4}$  
\\[20pt]

\hline\hline

\end{tabular*}
\end{sidewaystable}


\end{document}